\PassOptionsToPackage{sort, compress}{natbib}

\documentclass[11pt, a4paper]{gdm_format}
\usepackage[authoryear, sort&compress, round]{natbib}
\usepackage{listings}

\usepackage[most]{tcolorbox}
\usepackage{colortbl} 
\usepackage{xcolor} 
\usepackage{pdfpages}
\usepackage{booktabs}
\usepackage{adjustbox}
\usepackage{hyperref}
\usepackage{url}
\usepackage{booktabs}  
\usepackage{makecell}  
\usepackage{wrapfig}
\usepackage{multirow}  
\usepackage{array}  
\usepackage{booktabs} 
\usepackage{colortbl} 
\usepackage{xcolor} 
\usepackage{array} 
\usepackage{multirow} 
\usepackage{amssymb}
\usepackage{graphicx} 
\usepackage{float}    
\usepackage[most]{tcolorbox}
\usepackage{listings}
\usepackage{fontawesome5}
\usepackage{pgffor}
\usepackage[symbol]{footmisc}

\usepackage{amsmath}
\usepackage{amssymb} 
\usepackage{xspace}
\usepackage{multirow}
\usepackage{booktabs}
\usepackage{color}
\usepackage{colortbl}
\usepackage{cleveref}
\usepackage{graphicx}
\usepackage{algorithm}
\usepackage{algorithmicx}
\usepackage{algpseudocode}
\usepackage{subcaption}
\usepackage{wrapfig}
\usepackage{sidecap}
\usepackage{soul}
\usepackage{enumitem}
\sidecaptionvpos{figure}{t}
\usepackage{multicol}
\usepackage{pifont}
\usepackage[normalem]{ulem}
\usepackage{titletoc}

\usepackage{CJKutf8}

\lstdefinestyle{pythonstyle}{
    language=Python,
    basicstyle=\ttfamily\small,
    keywordstyle=\color{blue},
    commentstyle=\color{green!50!black},
    stringstyle=\color{red},
    showstringspaces=false,
    numbers=left,
    numberstyle=\tiny\color{gray},
    frame=single,
    breaklines=true,
    tabsize=4,
}

\newcommand{\blank}[1]{\hspace*{#1}\linebreak[0]}
\tcbset {
  base/.style={
    arc=0mm, 
    bottomtitle=-0.25mm,
    boxrule=0mm,
    colbacktitle=black!10!white, 
    coltitle=black, 
    fonttitle=\bfseries, 
    left=2.5mm,
    leftrule=1mm,
    right=3.5mm,
    title={#1},
    toptitle=0.25mm,
    breakable,
  }
}

\definecolor{brandblue}{rgb}{0.34, 0.7, 1}
\newtcolorbox{mybox}[1]{
  colframe=brandblue, 
  base={#1}
}

\definecolor{pink}{rgb}{1, 0.75, 0.8}
\newtcolorbox{safetybox}[1]{
  colframe=pink, 
  base={#1}
}

\title{Agent Laboratory: Using LLM Agents as Research Assistants}

\correspondingauthor{Samuel Schmidgall (sschmi46@jhu.edu)}

\author[1, 2]{Samuel Schmidgall}
\author[1]{Yusheng Su}
\author[1]{Ze Wang}
\author[1]{Ximeng Sun}
\author[1]{Jialian Wu}
\author[1]{Xiaodong Yu}
\author[1]{Jiang Liu}
\author[3]{Michael Moor}
\author[1]{Zicheng Liu}
\author[1]{Emad Barsoum}

\affil[1]{AMD}
\affil[2]{Johns Hopkins University}
\affil[3]{ETH Zurich}

\begin{document}

\begin{abstract}

Historically, scientific discovery has been a lengthy and costly process, demanding substantial time and resources from initial conception to final results.
To accelerate scientific discovery, reduce research costs, and improve research quality, we introduce \texttt{Agent Laboratory}, an autonomous LLM-based framework capable of completing the entire research process.
This framework accepts a human-provided research idea and progresses through three stages—literature review, experimentation, and report writing to produce comprehensive research outputs, including a code repository and a research report, while enabling users to provide feedback and guidance at each stage.
We deploy \texttt{Agent Laboratory} with various state-of-the-art LLMs and invite multiple researchers to assess its quality by participating in a survey, providing human feedback to guide the research process, and then evaluate the final paper. We found that: (1) \texttt{Agent Laboratory} driven by o1-preview generates the best research outcomes; (2) The generated machine learning code is able to achieve state-of-the-art performance compared to existing methods; (3) Human involvement, providing feedback at each stage, significantly improves the overall quality of research; (4) \texttt{Agent Laboratory} significantly reduces research expenses, achieving an 84\% decrease compared to previous autonomous research methods.
We hope \texttt{Agent Laboratory} enables researchers to allocate more effort toward creative ideation rather than low-level coding and writing, ultimately accelerating scientific discovery.

\end{abstract}

\maketitle

\begin{center}
\href{https://AgentLaboratory.github.io/}{\faGithub  \xspace \texttt{https://AgentLaboratory.github.io}}
\end{center}

\begin{figure}[ht!]
    \centering
    \includegraphics[width=0.93\linewidth]{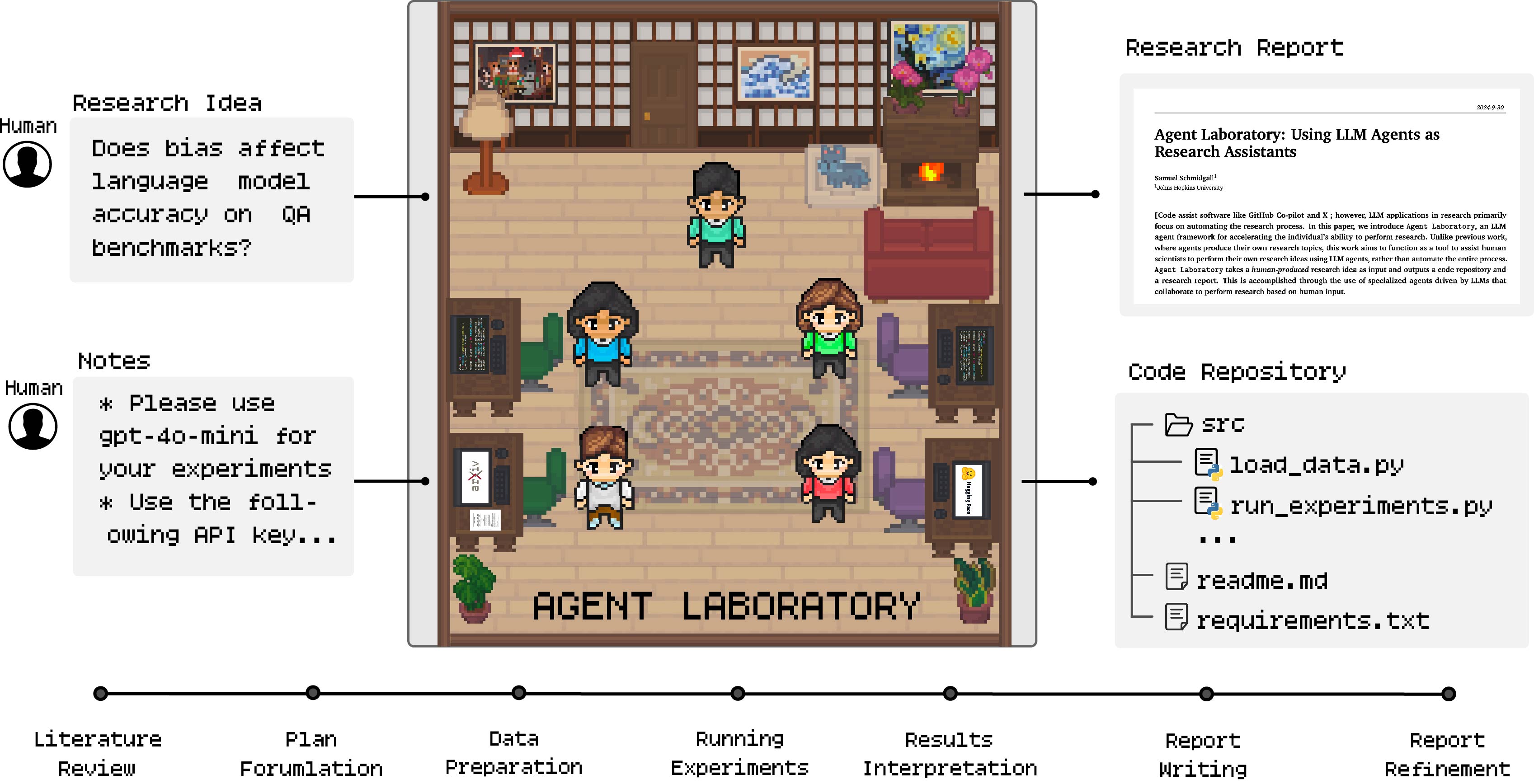}
    \caption{ \texttt{Agent Laboratory} takes as input a human research idea and a set of notes, provides this to a pipeline of specialized LLM-driven agents, and produces a research report and code repository.}
    \label{fig:AgentLab}
\end{figure}

\section{Introduction}

Scientists frequently face constraints that limit the number of research ideas they can explore at any given time, resulting in ideas being prioritized based on predicted impact. While this process helps determine which concepts are worth investing time in and how best to allocate limited resources effectively, many high quality ideas remain unexplored. If the process of exploring ideas had less limitations, researchers would be able to investigate multiple concepts simultaneously, increasing the likelihood of scientific discovery.

In an effort to achieve this, recent work has explored the capability of LLMs to perform research ideation and automated paper generation, where LLM agents perform the role of human scientists (\cite{baek2024researchagent, lu2024ai, ghafarollahi2024sciagents, swanson2024virtual}). The work of \cite{baek2024researchagent} introduces ResearchAgent, which automatically generates research ideas, methods, and experiment designs, iteratively refining them through feedback from multiple reviewing agents that mirror peer discussions and leverage human-aligned evaluation criteria to improve the outputs. \cite{lu2024ai} explores fully automated paper generation, where The AI Scientist framework generates novel research ideas, writes code, conducts experiments, and creates a full scientific paper with an automated peer-review system to evaluate the work. Even though these works demonstrate that current LLMs can generate ideas judged to be more novel than those produced by human experts, \cite{si2024can} indicates that LLMs still exhibit weaknesses in feasibility and implementation details, suggesting a complementary rather than replacement role for LLMs in research. Therefore, we aim to design an autonomous agent pipeline that can assist humans toward implementing their own research ideas.

In this work, we introduce \texttt{Agent Laboratory}, an autonomous pipeline for accelerating the individual’s ability to perform machine learning research. Unlike previous approaches, where agents participate in their own research ideation independent of human input (\cite{lu2024aiscientist, baek2024researchagent}), \texttt{Agent Laboratory} is designed to assist human scientists in executing their own research ideas using language agents. \texttt{Agent Laboratory} takes as input a human research idea and outputs a research report and code repository produced by autonomous language agents, allowing various levels of human involvement, where feedback can be provided at a frequency based on user preference. A detailed list of our contributions are provided below:

\begin{enumerate}
    \item We introduce \texttt{Agent Laboratory}, an open-source LLM agent framework for accelerating the individual’s ability to perform research in machine learning. In order to accommodate all users, \texttt{Agent Laboratory} is compute flexible, where various levels of compute can be allocated based on the individual's access to compute resource (e.g., CPU, GPU, memory) and model inference budget.
    \item Human evaluators rated papers generated using \texttt{Agent Laboratory} across experimental quality, report quality, and usefulness, showing that while the o1-preview backend was perceived as the most useful, o1-mini achieved the highest experimental quality scores, and gpt-4o was behind in all metrics.
    \item NeurIPS-style evaluations showed that o1-preview performed best among backends, particularly in clarity and soundness, according to human reviewers. However, a clear gap emerged between human and automated evaluations, with automated scores significantly overestimating quality (6.1/10 vs. 3.8/10 overall). Similar discrepancies were seen across clarity and contribution metrics, suggesting the need for human feedback to complement automated evaluations for more accurate assessments of research quality.
    \item Co-pilot mode in Agent Laboratory was evaluated on custom and preselected topics, showing higher overall scores compared to autonomous mode. Co-pilot  papers also saw trade-offs in experimental quality and usefulness, reflecting challenges in aligning agent outputs with researcher intent.
    \item The co-pilot feature in \texttt{Agent Laboratory} is overall found to have high utility and usability when rated by human users, with most participants deciding to continue usage after their experience
    \item Detailed cost and inference time statistics, as well as the breakdown of cost per paper phase, are presented for different model back-ends, demonstrating that \texttt{Agent Laboratory} offers automatic research at a greatly reduced price compared with other works (only \$2.33 USD per paper with a gpt-4o backend).
    \item State-of-the-art performance on a subset of MLE-Bench challenges using the proposed \texttt{mle-solver}, achieving higher consistency and scoring compared to other solvers, and earning more medals, including gold and silver, than MLAB, OpenHands, and AIDE.
\end{enumerate}

We hope that this work takes a step toward accelerating scientific discovery in machine learning, allowing researchers to allocate more effort toward creative ideation and experiment design rather than low-level coding and writing.

\section{Background \& Related Work}

\paragraph{Large language models} 

The research agents in this paper are built on autoregressive large language models (LLMs), which are trained on extensive text corpora to predict conditional probabilities of token sequences, \(p(x_t | x_{<t}; \theta)\), and generate text completions through sampling, where \(x_t \sim \text{softmax}(W \cdot h_t)\), with \(h_t\) as the hidden state and \(W\) as the learned weight matrix mapping to token probabilities. LLMs utilize transformer architectures (\cite{vaswani2017attention}) to capture long-range dependencies in text. These models, such as Claude (\cite{anthropic2024claude}), Llama (\cite{touvron2023llama1,touvron2023llama,dubey2024llama}), and ChatGPT (\cite{hurst2024gpt, openai_gpt3.5, achiam2023gpt}), leverage vast datasets and scaling techniques, thus enabling them to perform a wide array of language-based tasks, such as translation, summarization, and reasoning, by generalizing patterns learned during pretraining to novel inputs \cite{brown2020language}. 

\paragraph{LLM Agents}
While LLMs demonstrate strong understanding and reasoning abilities, they face challenges when executing tasks in real-world scenarios.
To overcome these limitations, their capabilities are extended through structured frameworks, enabling them to autonomously and semi-autonomously perform task execution and semi-autonomously perform task execution (\cite{wu2023autogen, li2023camel, chen2023agentverse, qian2024chatdev}).
These systems, referred to as agents, utilize techniques such as chain-of-thought prompting (\cite{wei2022chain}), iterative refinement (\cite{shinn2024reflexion}), self-improvement (\cite{huang2022large}), and external tool integration to execute complex workflows (\cite{hao2024toolkengpt, qin2023toolllm, schick2023toolformer}). 
LLM agents have made remarkable progress in solving tasks of real-world significance, such as software engineering (\cite{jimenez2023swe, yang2024swe, wang2024opendevin}), cybersecurity (\cite{abramovich2024enigma, wan2024cyberseceval, fang2024llm}), and medical diagnosis (\cite{tu2024towards, schmidgall2024agentclinic, mcduff2023towards}). There has also been progress in applying LLMs agents to embodied problems such as autonomous robotics (\cite{brohan2022rt, kimsurgical, black2024pi_0, brohan2023rt}), web tasks (\cite{gur2023real, putta2024agent, deng2024mind2web, shi2017world, he2024webvoyager}), and game playing (\cite{wang2023voyager, feng2024chessgpt, al2024project}).  For a broader overview of LLM agents, refer to \cite{wang2024survey}.

\paragraph{Automated machine learning}

Automated machine learning is an area of active research, with many approaches focused on using Kaggle, an online platform for machine learning competitions, as a benchmark for evaluating agent performance. Notable efforts include MLE-Bench (\cite{chan2024mle}), DS-bench (\cite{jing2024dsbench}), and MLAgentBench (\cite{huang2024mlagentbench}) which propose using 75, 74, and 6 Kaggle challenges respectively as benchmarks to measure the abilities of ML agents in tasks such as data preparation, model development, and submission.  Several ML "solvers" which can solve ML challenges have been introduced, such as AIDE (\cite{AIDE}), CodeActAgent (referred to as “OpenHands") (\cite{wang2024opendevin}), and ResearchAgent (referred to as “MLAB") from MLAgentBench (\cite{huang2024mlagentbench}) which automate feature implementation, bug fixing, and code refactoring with a high success rate.
Agent K (\cite{grosnit2024large}) demonstrates the ability to solve Kaggle challenges at the human-level with a challenge URL provided as input.

\paragraph{AI in Scientific Discovery}

AI has been used to support scientific discovery across numerous disciplines for decades. For instance, AI has been used for discovery in mathematics (\cite{romera2024mathematical}), material science (\cite{szymanski2023autonomous, pyzer2022accelerating, merchant2023scaling}), chemistry (\cite{jumper2021highly, hayes2024simulating}), algorithm discovery (\cite{fawzi2022discovering}), and computational biology (\cite{ding2024automating}). These approaches position AI as a tool rather than an agent performing research in autonomous research.

\paragraph{LLMs for research related tasks}

LLMs have demonstrated strong capabilities in diverse research-related tasks, such as code generation (\cite{chen2021evaluating, nijkamp2022codegen}), end-to-end software development (\cite{qian2024chatdev, qian2023experiential, phan2024hyperagent, hai2024repoexec}), code generation for discovery (\cite{majumder2024discoverybench, ifargan2024autonomous, hu2024infiagent, guo2024ds, gu2024blade, ghafarollahi2024protagents, chen2024scienceagentbench}), research question-answering (\cite{chen2024scholarchemqa, lala2023paperqa, song2024cs, lin2024biokgbench}), research ideation (\cite{baek2024researchagent, ghafarollahi2024sciagents, li2024chain, si2024can}), automated paper reviewing (\cite{d2024marg, liang2024can, lu2024aiscientist, weng2024cycleresearcher}), literature search (\cite{ajith2024litsearch, kang2024researcharena, press2024citeme, li2024scilitllm}), and predicting the outcome of experiments (\cite{luo2024large, ashokkumar2024predicting, lehr2024chatgpt, manning2024automated, zhang2024massw}). 
Although LLMs have made notable progress in solving the aforementioned tasks, ideation has struggled to progress, with some work showing that LLM ideation leads to greater novelty than humans (\cite{si2024can}), while others show reduced creativity (\cite{chakrabarty2024art}) and greater homogeneous effects (\cite{anderson2024homogenization, zhou2024shared}) that may limit creative discovery without human guidance.

Additionally, research on human-AI collaboration has reached mixed conclusions about the idea novelty (\cite{ashkinaze2024ai, liu2024ai, padmakumardoes}). These findings suggest that, with the current LLMs, the strongest research systems would combine human-guided ideation with LLM-based workflows.

\paragraph{LLMs for autonomous research}

Recent advancements in automated scientific workflows have focused on leveraging LLMs to emulate the process of research. \cite{swanson2024virtual} introduces a team of LLM agents working as scientists alongside a human researcher who provides high-level feedback, with the end result being novel nanobody binders aimed at addressing recent variants of SARS-CoV-2. ChemCrow (\cite{ m2024augmenting}) and Coscientist (\cite{boiko2023autonomous}) demonstrate the ability for autonomous ideation and experimentation in chemistry. ResearchAgent (\cite{baek2024researchagent}) automates research idea generation, experiment design, and iterative refinement using feedback from reviewing agents aligned with human evaluation criterion. 
The AI Scientist (\cite{lu2024ai}) extends this automation to encompass end-to-end scientific discovery, including coding, experiment execution, and automated peer review for manuscript generation. Despite these advancements, studies like \cite{si2024can} highlight limitations in the feasibility and implementation details of LLM ideation, indicating a complementary rather than replacement role for LLMs in autonomous research.

\section{Agent Laboratory}

\begin{figure}
    \centering
    \includegraphics[width=0.99\linewidth]{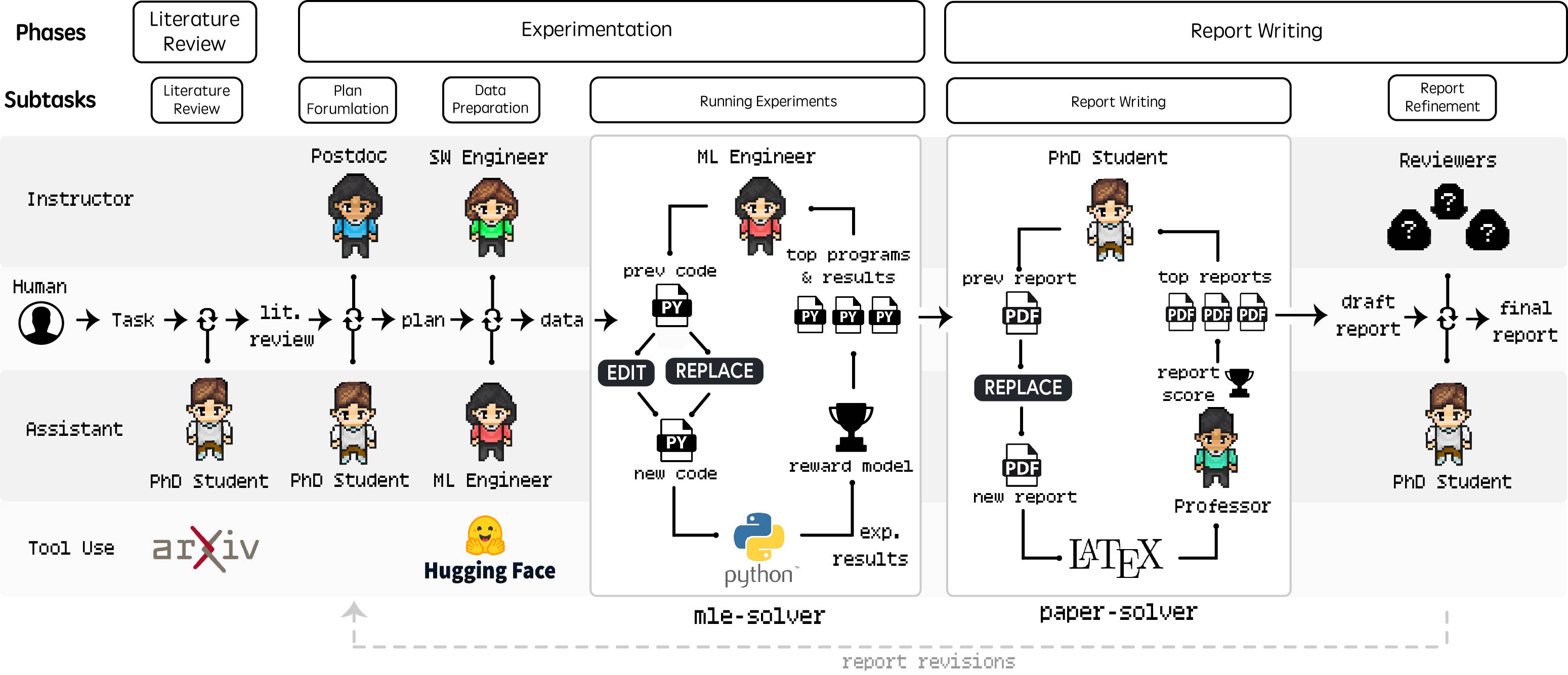}
    \caption{\texttt{Agent Laboratory} Workflow. This image illustrates the three primary phases of Agent Laboratory: Literature Review, Experimentation, and Report Writing, each featuring distinct tasks, tools, and human-agent roles. The pipeline integrates human input with LLM-driven agents, such as the PhD and Postdoc agents, which handle literature reviews, experimental planning, data preparation, and result interpretation. Specialized tools like mle-solver for experimentation and paper-solver for report generation automate tedious research tasks, enabling collaboration between human researchers and AI to produce high-quality research outputs.}
    \label{fig:AgentWorkflow}
\end{figure}

\paragraph{Overview.} \texttt{Agent Laboratory} begins with the independent collection and analysis of relevant research papers, progresses through collaborative planning and data preparation, and results in automated experimentation and comprehensive report generation. 
As shown in Figure \ref{fig:AgentWorkflow}, the overall workflow consists of three primary phases: (1) Literature Review, (2) Experimentation, and (3) Report Writing. 
In this section, we will introduce these phases in detail along with the corresponding involved agents.
Furthermore, in Section \ref{sec:results}, we will conduct qualitative and quantitative analyses to demonstrate the strengths of \texttt{Agent Laboratory} and its ability to generate research.

\subsection{Literature Review}
\label{sec:lit_review}
\paragraph{Literature Review.} 
The literature review phase involves gathering and curating relevant research papers for the given research idea to provide references for subsequent stages. During this process, the PhD agent utilizes the arXiv API to retrieve related papers and performs three main actions: \texttt{summary}, \texttt{full text}, and \texttt{add paper}. The \texttt{summary} action retrieves abstracts of the top 20 papers relevant to the initial query produced by the agent. The \texttt{full text} action extracts the complete content of specific papers, and the \texttt{add paper} action incorporates selected summaries or full texts into the curated review. This process is iterative rather than a single-step operation, as the agent performs multiple queries, evaluates the relevance of each paper based on its content, and refines the selection to build a comprehensive review. Once the specified number of relevant texts (N=max) is reached via the \texttt{add paper} command, the curated review is finalized for use in subsequent phases.

\subsection{Experimentation}
\label{sec:experimentation}
\paragraph{Plan Formulation} The plan formulation phase focuses on creating a detailed, actionable research plan based on the literature review and research goal. During this phase, the PhD and Postdoc agents collaborate through dialogue to specify how to achieve the research objective, detailing experimental components needed to complete the specified research idea such as which machine learning models to implement, which datasets to use, and the high-level steps of the experiment. Once a consensus is reached, the Postdoc agent submits this plan using the \texttt{plan} command, which serves as a set of instructions for subsequent subtasks.

\paragraph{Data Preparation.} The goal of the data preparation phase is to write code that prepares data for running experiments, using the instructions from the plan formulation stage as a guideline. The ML Engineer agent executes code using \texttt{Python} command command and observes any printed output. The ML Engineer has access to HuggingFace datasets, searchable via the \texttt{search HF} command. After agreeing on the finalized data preparation code, the SW Engineer agent submits it using the \texttt{submit code} command. Before the final submission proceeds, the code is first passed through a Python compiler to ensure that there are no compilation issues. This process will be iteratively executed until the code is bug-free.

\paragraph{Running Experiments.} In the running experiments phase, the ML Engineer agent focuses on implementing and executing the experimental plan formulated prior. This is facilitated by \texttt{mle-solver}, a specialized module designed to generate, test, and refine machine learning code autonomously. \texttt{mle-solver} begins by producing initial code based on the research plan and insights from the literature review. For the first \texttt{mle-solver} step, the program is empty and must generate a file from scratch, which is used as the \textit{top scoring program}. The following processes describe the workflow of the \texttt{mle-solver}:


\begin{figure}
    \centering
    \includegraphics[width=0.90\linewidth]{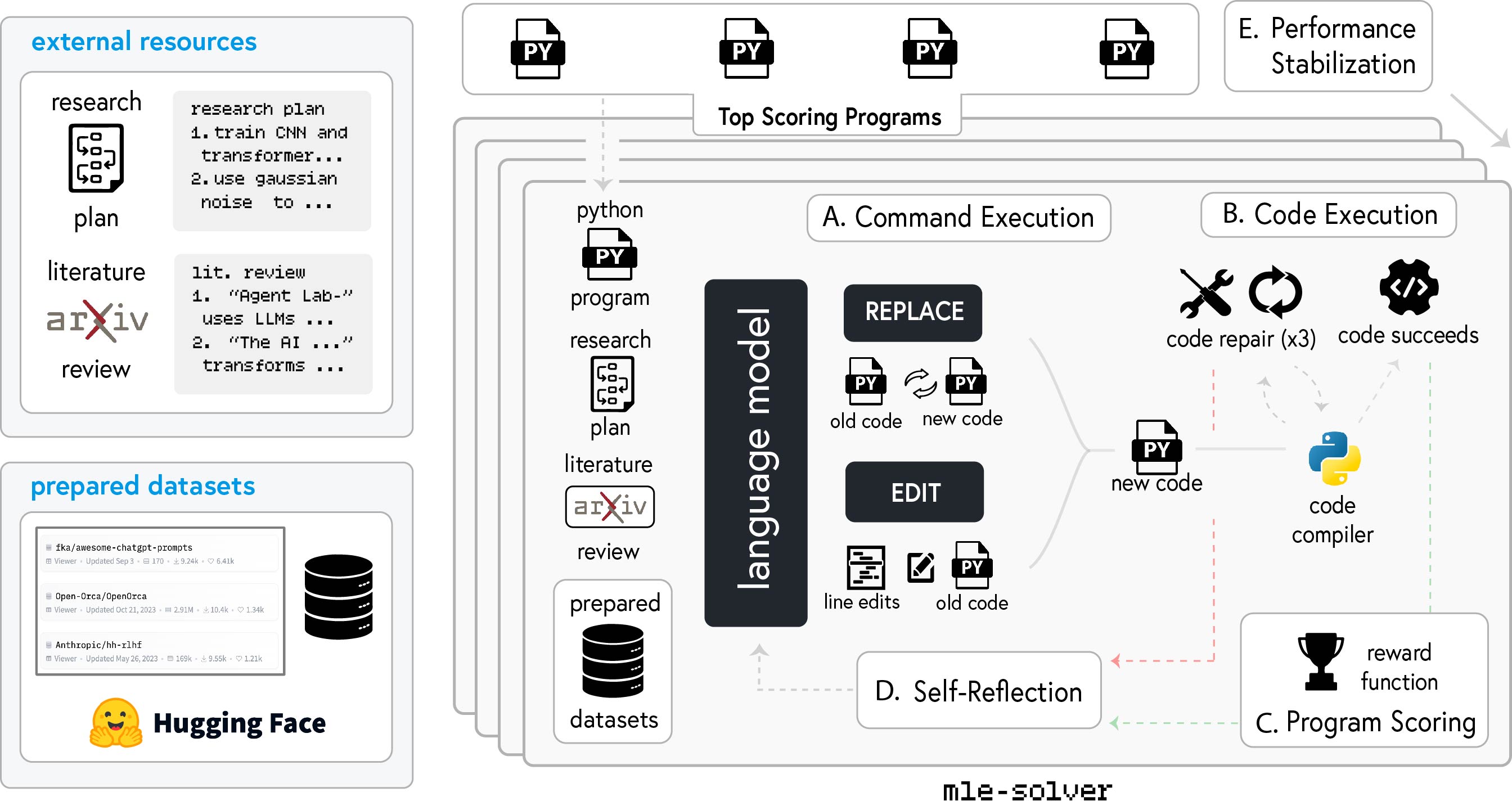}
    \caption{Overview of the \texttt{mle-solver} workflow. This diagram details the iterative process used by the MLE-Solver to autonomously generate  machine learning code. Beginning with external resources, the workflow integrates command execution (A), where new code is generated, followed by code execution (B) to compile and repair issues if needed. Program scoring (C) evaluates the generated code using a reward function, while self-reflection (D) helps refine future iterations based on results. Performance stabilization (E) ensures consistent outcomes by maintaining a pool of top-performing programs and iterative optimization.}
    \label{fig:mle-solver}
\end{figure}

\begin{enumerate}
    \item [A.] \textbf{Command Execution.} During the command execution phase, an initial program is sampled from a maintained set of top-performing programs, which is represented by a single file during initialization. The \texttt{mle-solver} iteratively refines this program through two operations, \texttt{REPLACE} and \texttt{EDIT}, to better align the output with experimental objectives. The \texttt{EDIT} operation identifies a range of lines, substituting the code between the specified line numbers with newly generated code. In contrast, the \texttt{REPLACE} operation generates a completely new Python file.
    \item [B.] \textbf{Code Execution.} After a code command is executed, the new program is passed through a compiler to check for runtime errors. If it successfully compiles, a score is returned and the list of top programs is updated if the score is higher than the existing programs. If the code does not compile, the agent attempts to repair the code for $N_{rep}$ tries ($N_{rep}$=3 in our experiments) before returning an error and moving on to a new code replacement.
    \item [C.] \textbf{Program Scoring.} If a code succeeds in compilation, it is sent to a scoring function which determines if it is better than previously implemented experiment code. In order to obtain a program score, we implement a scoring function that uses an LLM reward model to assess the effectiveness of the ML code generated by \texttt{mle-solver}. The reward model, invoked as an LM, scores the program on a scale from 0 to 1 considering the outlined research plan, the produced code, and the observed output to determine how accurately the program adheres to the initial goals. A score of 1 is provided for results with high alignment and everything below on a spectrum of how closely the output and code matches the planning goals. This process is similar to existing methods for LLM reasoning tree search (\cite{yao2024tree}), where instead of a series of reasoning steps being traversed using self-evaluated LLM scoring, the set of possible programs are being traversed (via \texttt{EDIT} and \texttt{REPLACE} commands) and the resulting program outcome is self-evaluated to determine if a program is worth building on. This is similar to the Solution Space Search of AIDE (\cite{AIDE}), however their method was specifically designed for the Kaggle competitions and is simply  extracting the accuracy rather than scoring the research code and outcomes.
    \item [D.] \textbf{Self Reflection.} Whether the code succeeds or fails, a self-reflection is produced based on the experimental results or the encountered error signal (\cite{shinn2024reflexion,renze2024self}). Here, the \texttt{mle-solver} is prompted to reflect on the outcome of its actions. If the program failed to compile, the solver reflects on how to fix this issue in next iterations. If it successfuly compiles and returns a score, the solver will reflect on how to increase this score. These reflections are generated to improve future performance, ensuring that the system learns from errors, improving the quality and robustness of the generated code over iterative cycles.
    \item [E.] \textbf{Performance Stabilization} To prevent performance drift, two mechanisms are implemented: top program sampling and batch-parallelization. In top program sampling, a collection of the highest-scoring programs is maintained, and one program is randomly sampled before executing a command, ensuring diversity while retaining quality. For batch-parallelization, each solver step involves making N modifications simultaneously, with the top modification selected to replace the lowest-scoring program in the top collection. These strategies use high-entropy sampling to modify the code, resulting in a balance between exploration of new solutions and refinement of existing ones in order to maintain stable code modifications. 
\end{enumerate}

\paragraph{Results Interpretation.} The goal of the results interpretation phase is to derive meaningful insights from experimental outcomes to inform the final report. The PhD and Postdoc agents discuss their understanding of the experimental results produced by \texttt{mle-solver}. Once they agree on a meaningful interpretation that could contribute to a compelling academic paper, the Postdoc agent submits it using the \texttt{interpretation} command, forming the basis for the report writing phase.

\subsection{Report Writing} 
\label{sec:report_writing}

\paragraph{Report Writing.} In the report writing phase, the PhD and Professor agent synthesize the research findings into a comprehensive academic report. This process is facilitated by a specialized module called \texttt{paper-solver}, which iteratively generates and refines the report. The \texttt{paper-solver} aims to act as a report generator, positioning the work that has been produced by previous stages of \texttt{Agent Laboratory}. \texttt{paper-solver} does not aim to entirely replace the academic paper-writing process, but rather to summarize the research that has been produced in a human-readable format so that the researcher \textit{using} \texttt{Agent Laboratory} understands what has been accomplished.  
The output follows the standard structure of an academic paper, ensuring it meets conference submission requirements (for the paper scoring phase) while being clear and methodical. The following processes describe the workflow of \texttt{paper-solver}:

\begin{figure}
    \centering
    \includegraphics[width=0.99\linewidth]{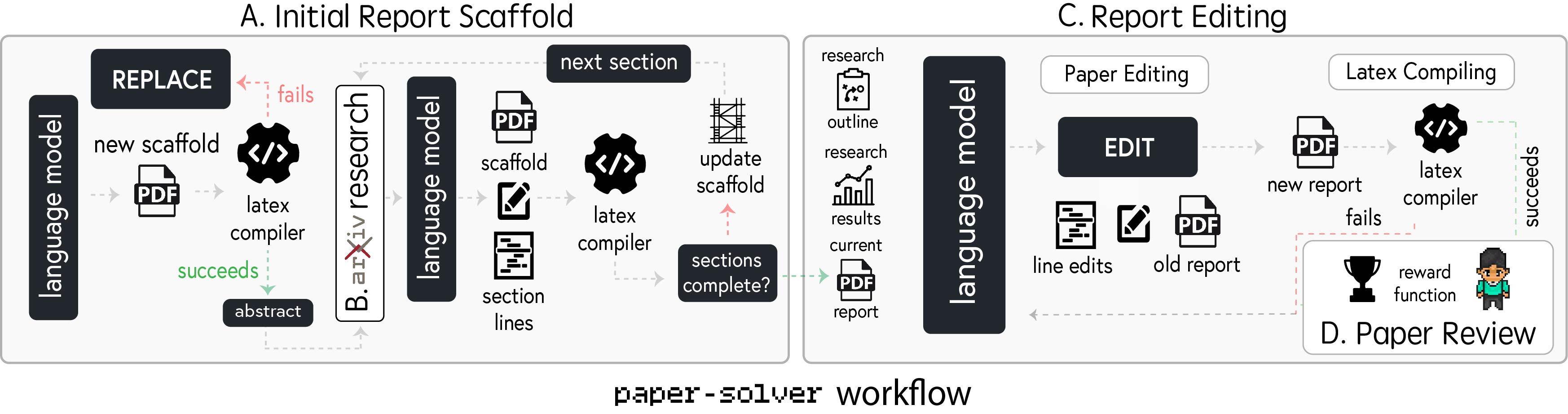}
    \caption{Graphical outline of \texttt{paper-solver}. This diagram showcases the step-by-step process of generating and refining academic research reports using the Paper-Solver tool. The workflow starts with the creation of an initial report scaffold (A) by iteratively generating LaTeX-based sections, followed by updates to ensure structural completeness. (B) Research is performed through an Arxiv tool during relevant sections. In the Report Editing phase (C), the language model applies targeted edits to improve the document, with LaTeX compilation verifying the integrity of changes. Finally, the completed report undergoes a reward-based evaluation during the Paper Review phase (D), ensuring alignment with academic standards and research goals.}
    \label{fig:PaperSolver}
\end{figure}

\begin{enumerate}
    \item [A.] \textbf{Initial Report Scaffold.} The first task of the \texttt{paper-solver} is to generate an initial scaffold for the research paper. This scaffold outlines the document structure, dividing it into eight standardized sections: Abstract, Introduction, Background, Related Work, Methods, Experimental Setup, Results, and Discussion. During scaffold creation, placeholders are inserted for each section to categorize future content. This process establishes the framework for subsequent detailed text generation. The scaffold includes necessary formatting for LaTeX compilation, allowing the generated paper to be directly reviewed and refined. Special care is taken to ensure the scaffold aligns with academic conventions, such as appropriate section titles and placeholders that guide content development.
    \item [B.] \textbf{Arxiv Research.} During the scaffold building phase, we allow the \texttt{paper-solver} access to arXiv which is accessible through the same interface as the earlier literature review phase. ArXiv is enabled to allow the solver to explore related literature on the subject it is writing on as well as finding papers to refer to, although it is not enforced. We note that the agent still has access to the original literature search, but has the opportunity to expand based on literature needed to write a particular paper section.
    \item [C.] \textbf{Report Editing.} One the scaffold is built, the \texttt{paper-solver} uses specialized commands to iteratively refine the generated paper. The primary command are available for this stage is the \texttt{EDIT} command, which allows precise line-by-line modifications to the LaTeX code. This command enable dynamic adjustments to the content, ensuring alignment with the research plan, the clarity of arguments, and compliance with formatting standards. Before integrating edits, the system compiles the LaTeX to verify error-free functionality, thereby maintaining document integrity. Through iterative editing, the solver ensures the paper achieves the desired level of quality, cohesiveness, and depth required for academic acceptance. 
    \item [D.] \textbf{Paper Review.} For obtaining scores for papers during the \texttt{paper-solver} iterations, we leverage an adapted version of the automated review system developed in \cite{lu2024aiscientist}. This system works by using an LLM-based agent to simulate the scientific paper review process following the NeurIPS conference guidelines.
    When evaluated on 500 ICLR 2022 papers from the OpenReview dataset, the automated reviewer achieved human-level accuracy (65\% compared to 66\% for human reviewers) and surpassed human performance in F1 score (0.57 vs. 0.49) after calibration. 
    An example review from one of our papers by o1-mini is provided below.
\end{enumerate}

\begin{tcolorbox}[breakable,colback=blue!5!white, colframe=blue!70!black, title=Example Review ( o1-mini | Word Order Sensitivity )]
\texttt{
 "Strengths": \text{[}\\
	\blank{1cm} "Comprehensive experimental design and methodology.",\\
	\blank{1cm} "Use of a well-known dataset (RACE) for evaluation.",\\
	\blank{1cm} "Empirical validation of bias mitigation strategies.",\\
	\blank{1cm} "Clear presentation of results and analysis."\text{]},\\
Weaknesses": \text{[}\\
	\blank{1cm} "Limited exploration of additional bias mitigation techniques.",
	\blank{1cm} "Lack of in-depth discussion on limitations \\\blank{1cm} and societal impacts.",\\
	\blank{1cm} "The originality could be enhanced by exploring novel \\\blank{1cm} strategies."\text{]},\\
"Originality": 3, "Quality": 4, "Clarity": 3, "Significance": 3,\\
"Questions": \text{[}\\
	\blank{1cm} "Have you considered exploring additional bias \\\blank{1cm} mitigation techniques beyond majority voting and entropy-based \\\blank{1cm} thresholding?",\\
	\blank{1cm} "Can you provide more details on the potential societal impacts\\\blank{1cm}  of the model's sensitivity to option order?",\\
	\blank{1cm} "What are the limitations of the current study, and how\\\blank{1cm}  might they be addressed in future work?"\text{]},\\
"Limitations": \text{[}\\
	\blank{1cm} "The study is limited to the RACE dataset and may not generalize\\\blank{1cm}  to other datasets.",\\
	\blank{1cm} "The bias mitigation strategies, while effective,\\\blank{1cm}  do not completely eliminate sensitivity to option order."\text{]},\\
	"Ethical Concerns": false,\\
	"Soundness": 3, "Presentation": 3, "Contribution": 3, \\"Overall": 7, "Confidence": 4, \\"Decision": "Accept"
}
\end{tcolorbox}

\paragraph{Paper Refinement.}  In the paper refinement phase, the PhD agent makes a decision on whether to make paper revisions or to determine that the paper is complete. The process begins with a set of three reviewer agents generating reviews that mimic feedback from NeurIPS peer reviewers, evaluating the report based on criteria such as originality, quality, clarity, and significance. Based on these scores, the PhD agent then decides whether to finalize the project or revisit earlier subtasks—such as planning, experimentation, or results interpretation—to address the feedback. This allows the agents to refine the research report until it meets sufficiently high standards, effectively simulating the real-world academic revision process.

\subsubsection{Autonomous versus Co-Pilot Mode:} 
\label{sec:auto_vs_copilot}

There are two ways in which \texttt{Agent Laboratory} can be operated: autonomous and co-pilot modes. In autonomous mode, there is no human involvement other than providing the initial research idea for agents to produce research for. Each subtask moves on to the next subtask sequentially upon completion. In co-pilot mode, in addition to providing the research idea, there is also a checkpoint at the end of each subtask, where a human is involved in reviewing the work produced by agents in that phase (e.g., the literature review summary or generated report). The human reviewer can either decide to proceed to the next subtask, or ask the agent to repeat the subtask while providing high level notes for the agent to improve its performance during the next attempt. For example, if the literature review phase did not include a specific paper or the experiments did not include a desired technique, the human reviewer would instruct the agent to include this.

\section{Results}
\label{sec:results}

In this section, we present our main findings on the efficacy of \texttt{Agent Laboratory} to produce research.
We begin our results by asking how human evaluators perceive papers generated by \texttt{Agent Laboratory} running in end-to-end autonomous mode across five topics. Next, we examine human evaluation when using \texttt{Agent Laboratory} in collaborative co-pilot mode from both allowing the researcher to choose any topic they want and from our set of preselected topics. We then provide a detailed runtime analysis including cost, average time, and success rate by various models. Finally, we conclude with an evaluation of the \texttt{mle-solver} in isolation on MLE-Bench, a set of real-world Kaggle challenges. The details of all surveys are provided in Appendix \ref{appendix:surveys}.

\subsection{Evaluation of quality by language model}
\label{sec:humanpercept}

\begin{figure}
    \centering
    \includegraphics[width=0.99\linewidth]{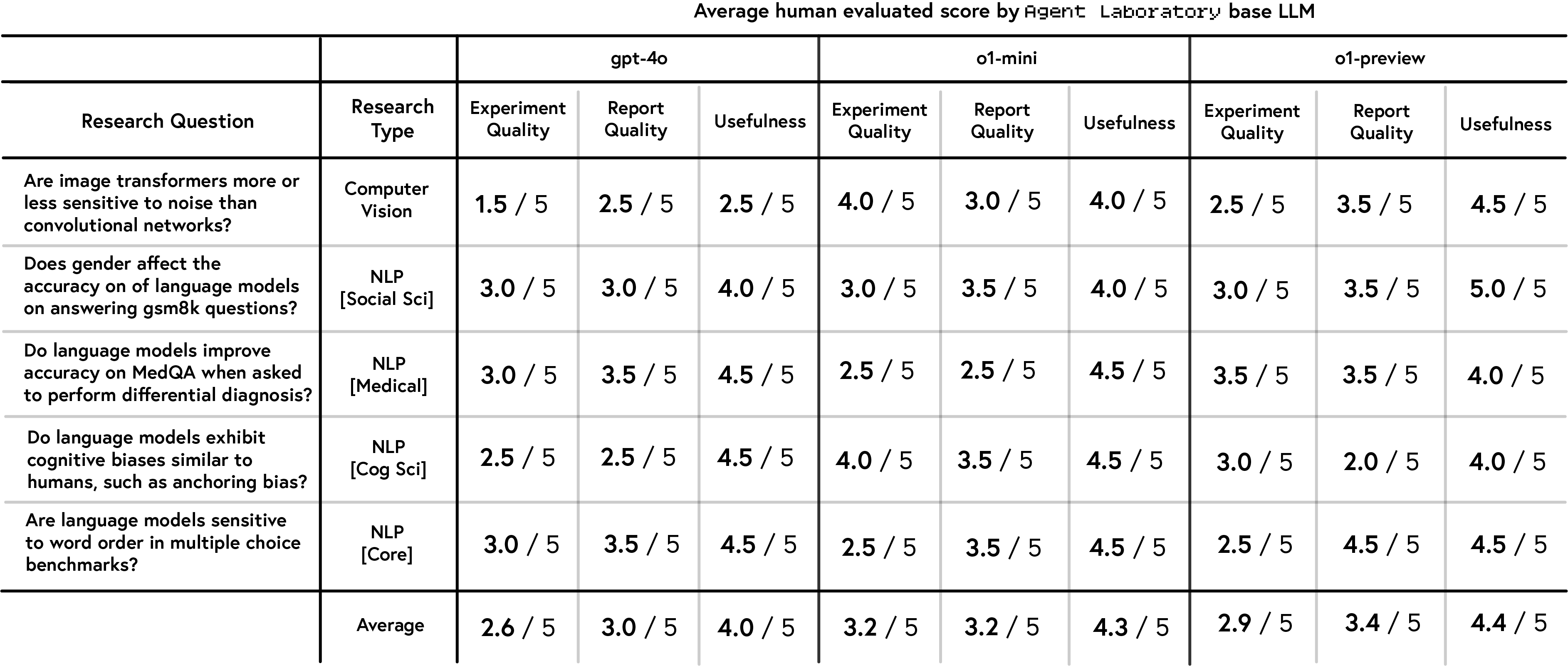}
    \caption{The average human evaluated scores of papers generated by \texttt{Agent Laboratory} in an autonomous mode based on a research question (left column) and LLM backend (top row). The bottom row shows the average score across all topics by LLM backend.}
    \label{fig:PaperScores}
\end{figure}

Our first experiment aims to evaluate how human-evaluated quality varies across three axes: experiment quality, report quality, and usefulness. This evaluation was conducted by human participants using three different LLM backends: gpt-4o (\cite{hurst2024gpt}), o1-mini, and o1-preview (\cite{openai2024introducing}). Research questions were selected from a set of 5 templates: 
\begin{enumerate}
    \item Do language models exhibit cognitive biases, such as confirmation bias or anchoring bias?
    \item Are image transformers more or less sensitive to pixel noise than convolutional networks?
    \item Do language models improve accuracy on MedQA when asked to perform differential diagnosis?
    \item Are language models sensitive to word order in multiple choice benchmarks?
    \item Does gender role play affect the accuracy on of language models on answering math questions?
\end{enumerate}

These 5 questions across 3 LLM backends resulted in a total of 15 papers being written autonomously by \texttt{Agent Laboratory} without any human involvement. We then recruited 10 volunteer PhD students to review 3 randomly assigned papers each. These researchers rated the experimental quality, report quality, and usefulness of the generated outputs on a scale of 1 to 5. The goal of this evaluation is to understand the differences in quality of produced research based on the three distinct LLM backbones, and to understand the usefulness of \texttt{Agent Laboratory} in autonomous mode. The details of the evaluation questions are provided here:

\begin{itemize}
    \item \textbf{Experimental Quality:} What is your perception of the quality of the experimental results presented in this report?
    \item \textbf{Report Quality:} What is your perception of the quality of the research report writing quality presented in this report?
    \item \textbf{Usefulness:} What is your perception of the usefulness of an AI assistant tool that can generate the presented report autonomously?
\end{itemize}

The results of this evaluation indicate variability in performance across different \texttt{Agent Laboratory} LLM backends (Figure \ref{fig:PaperScores}). gpt-4o consistently achieved lower scores, with an average experimental quality rating of 2.6/5, a report quality rating of 3.0/5, and a usefulness rating of 4.0/5. In contrast, o1-mini generally outperformed gpt-4o in experimental quality, with an average score of 3.2/5 (\textcolor{ForestGreen}{+0.6}), while maintaining similar levels of report quality and usefulness at 3.2/5 (\textcolor{ForestGreen}{+0.2}) and 4.3/5 (\textcolor{ForestGreen}{+0.3}), respectively. o1-preview demonstrated the highest usefulness and report quality, averaging 4.4/5 (\textcolor{ForestGreen}{+0.4} from gpt-4o and \textcolor{ForestGreen}{+0.1} from o1-mini) and 3.4/5 (\textcolor{ForestGreen}{+0.4} from gpt-4o and \textcolor{ForestGreen}{+0.2} from o1-mini) respectively, though its experimental ratings were slightly lower than o1-mini at 2.9/5 (\textcolor{ForestGreen}{+0.3} from gpt-4o and \textcolor{Magenta}{-0.3} from o1-mini). While all backends perform comparably in terms of report and experimental quality, the o1-preview model was as the most useful for research assistance, suggesting that its outputs were better aligned with the expectations and needs of researchers.

From our results, the quality is demonstrated to vary based on the selected topic. We find that the overall highest average report quality to be 3.8/5 and usefulness to be 4.5/5 for the \textit{word order} topic and the highest average experiment quality to be 3.2/5 for the \textit{cognitive bias} topic. Interestingly, we also find that \textit{word order} has the lowest experiment quality at 2.7/5 along with the \textit{image noise} topic. The \textit{image noise} topic was demonstrated to have high variance based on the LLM backend, with an experiment quality score of 1.5/5 for gpt-4o and a 4.0/5 with o1-mini (\textcolor{ForestGreen}{+2.5} point difference) and a usefulness score of 2.5/5 for gpt-4o and a 4.5/5 with o1-mini (\textcolor{ForestGreen}{+2.0} point difference).

In summary, the evaluation of quality across LLM backends demonstrates clear differences in experimental quality, report quality, and usefulness. While o1-preview is consistently rated as the most useful for research assistance, o1-mini achieves the highest experimental quality scores, and gpt-4o is generally being outperformed in all areas. Topic-specific trends suggest there may exist variability in the performance of \texttt{Agent Laboratory} across difference areas of machine learning research and across backend models.

\subsubsection{Human reviewer scores by language model}
\label{sec:humanrevscores}

\begin{figure}
    \centering
    \includegraphics[width=0.99\linewidth]{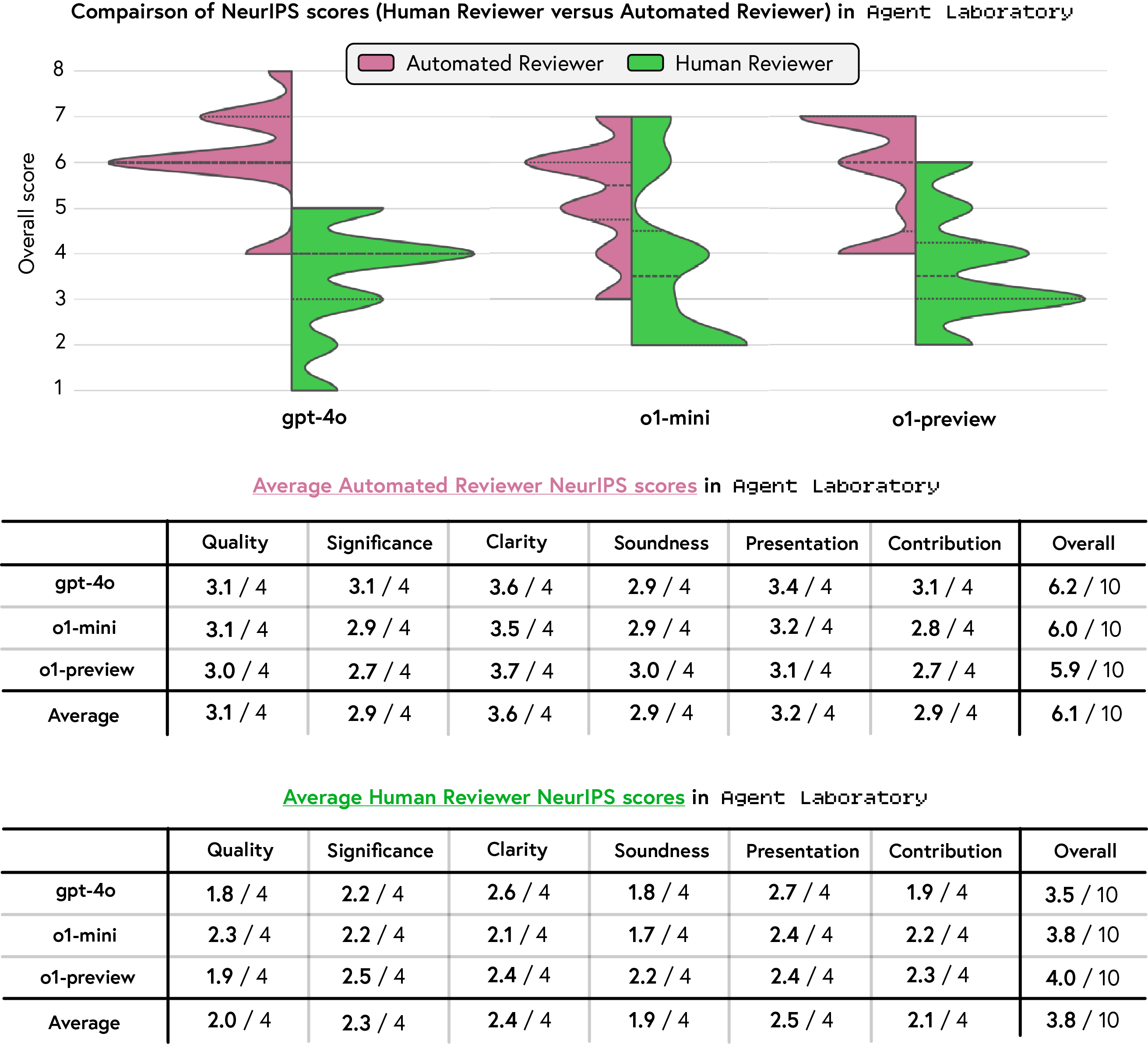}
    \caption{Scores from NeurIPs-style evaluation of generated papers, including the criterion: quality, significance, clarity, soundness, presentation, and contribution. (top) Split-violin plot comparing the overall score distribution of automated reviewers (LLM scores, left half of violin) and human reviewers (right half of violin). Human scores are not predictive of automated reviewer scores, demonstrating an average of \textcolor{Magenta}{-2.3} points lower. (middle) Automated reviewer scores across NeurIPs-style criterion. (bottom) Human reviewer scores across NeurIPs-style criterion.}
    \label{fig:humanscores}
\end{figure}

In addition to evaluating paper quality, we also asked human reviewers to assess papers generated by \texttt{Agent Laboratory} according to NeurIPS-style criteria, including quality, significance, clarity, soundness, presentation, and contribution as shown in Figure \ref{fig:humanscores}. We evaluated the same papers analyzed in Section \ref{sec:humanpercept} using the aforementioned metrics and conducted the comparison. We found that the average human scores for the three backends revealed differences in performance, with average overall ratings ranging from 3.5/10 with gpt-4o, 3.8/10 with o1-mini, and 4.0/10 with o1-preview.

First, when evaluating quality we find that reviewers rated gpt-4o the lowest at 1.8/4, while o1-mini achieved the highest score of 2.3/4, demonstrating relatively better technical soundness. In terms of significance, all three backends received similar scores between 2.2–2.5/4, indicating a modest contribution to advancing research goals. Clarity scores showed slight variability, with gpt-4o receiving 2.6/4 and o1-mini falling slightly lower at 2.1/4 (\textcolor{Magenta}{-0.5}), reflecting differences in how well the papers were written.
The soundness of the generated outputs, which assesses the robustness of claims, was rated highest for o1-preview at 2.2/4, with o1-mini and gpt-4o at 1.8 (\textcolor{Magenta}{-0.4}) and 1.7. Presentation and contribution ratings followed similar trends, with the overall contribution score averaging 2.1/4 across models, highlighting a need for improvement in the originality of the outputs.

These scores show a general trend where human reviewers identified o1-preview as producing slightly better-rounded outputs compared to other backends, though significant gaps remain in technical and methodological aspects across all models. We note that the average score of an accepted paper at NeurIPS is 5.9. In this regard, on average, papers produced in autonomous mode are below the acceptance threshold for top ML conferences. These results demonstrate that, in autonomous mode, there is a need for refinement of \texttt{Agent Laboratory} to meet human expectations for high-quality, impactful research papers.

\paragraph{Automated Reviews versus Human Reviews.} We also explore to what extent the automated reviewer scores align with those of human reviewers. The alignment is graphically illustrated using both tabular data (for all scores) and violin plots (for overall scores) in Figure \ref{fig:humanscores}. Our findings suggest that automated reviewers demonstrate notable discrepancies across all metrics compared with human evaluators, with a tendency to highly over-estimate the contribution of self-evaluated work. While the automated reviewers gave an average overall above average NeurIPS paper score of 6.1/10, human reviewers provided a much lower average of 3.8/10 (\textcolor{Magenta}{-2.3} points). Similar gaps are observed for all specific criteria, such as clarity and contribution, where automated reviewers rated clarity at 3.6/4 on average compared to 2.4/4 by human evaluators. This pattern holds for all criterion. Previous work demonstrates high alignment with automated reviewers (\cite{lu2024aiscientist}) and ICLR scores from OpenReview. However, with actual humans rating the generated papers, we find that automated reviews do not align closely with human reviews and are far from an average accepted paper at NeurIPS 2024, which stands at 5.85\footnote{\href{https://papercopilot.com/statistics/neurips-statistics/neurips-2024-statistics/}{https://papercopilot.com/statistics/neurips-statistics/neurips-2024-statistics}} (our scores were \textcolor{Magenta}{-2.05} points lower on average). Our results demonstrate that it is important for human evaluations to be provided alongside automated reviewer scores in future works in order to obtain a better understanding of the quality of generated papers.

\subsection{Evaluation of co-pilot quality}

We next evaluate the use of \texttt{Agent Laboratory} in co-pilot mode, where a human researcher is providing feedback at the end of each subtask (see Section \ref{sec:auto_vs_copilot} for more details). We evaluate performance across two measures: (1) the quality of \texttt{Agent Laboratory} as a tool for assisting their research and (2) the quality of generated papers. We first ask researchers to co-pilot \texttt{Agent Laboratory} on a topic of their choice without limitations. We then ask researchers to select a topic from the 5 topics introduced in Section \ref{sec:humanpercept}, resulting in a total of 2 papers per researcher which we refer to as \textbf{custom} and \textbf{preselected} papers respectively. After their papers are generated, we ask researchers to rate their experience using \texttt{Agent Laboratory} during the process of generating custom and preselected papers. We then ask them to self-evaluate the generated papers according to NeurIPS-style criterion. Finally, we ask external researchers to evaluate their paper comparing performance with \texttt{Agent Laboratory} in autonomous mode. All experiments used an o1-mini backbone for all phases except the literature review. 

\subsubsection{Quality as a tool}

The evaluation of \texttt{Agent Laboratory} as a research tool focuses on understanding its effectiveness in assisting researchers during the co-pilot mode. After generating their papers, participants were asked to reflect on their experiences and assess the tool's utility, usability, and overall satisfaction. We begin our evaluation by asking the following questions: 

\begin{itemize}
    \item \textbf{Utility:} How useful is \texttt{Agent Laboratory} for assisting your research?
    \item \textbf{Continuation:} How likely are you to continue using Agent Laboratory for research?
    \item \textbf{Satisfaction:} How much did you enjoy using Agent Laboratory?
    \item \textbf{Usability:} How easy was it for you to build a project using Agent Laboratory?
\end{itemize}

\begin{figure}
    \centering
    \includegraphics[width=0.99\linewidth]{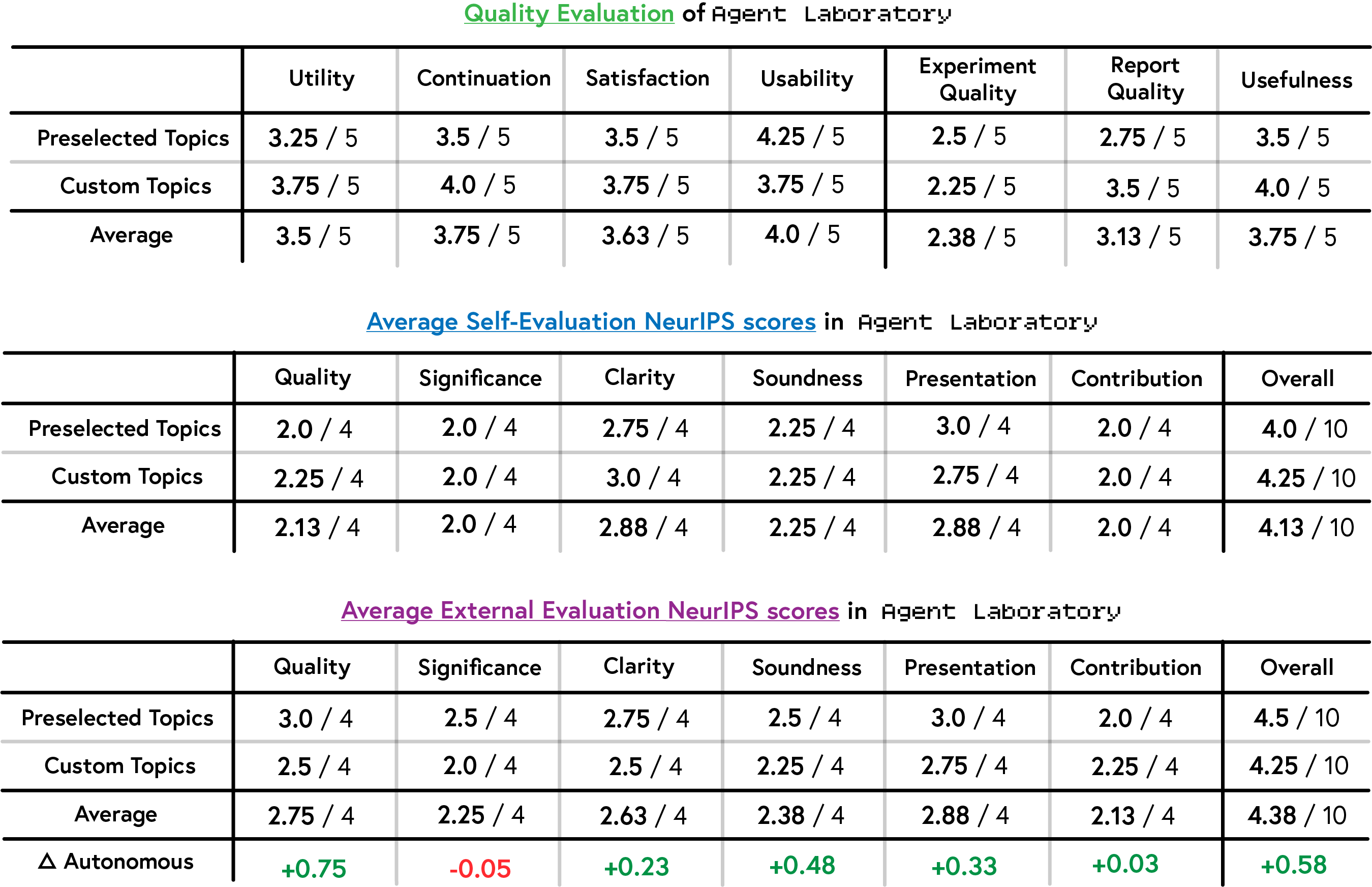}
    \caption{Co-pilot evaluation.}
    \label{fig:copilot_fig}
\end{figure}

The result of answering each question is a score from 1-5, where 1 indicates the lowest agreement and 5 indicates the highest. We find that the overall scores across all experiments are 3.5/5 for utility, 3.75/5 for continuation, 3.63/5 for satisfaction, and 4.0/5 for usability (Figure \ref{fig:copilot_fig}).
We also delineate average scores based on custom and preselected topics. For custom experiments, we find overall scores of 3.75/5 for utility, 4.0/5 for continuation, 3.75/5 for satisfaction, and 3.75/5 for usability. For preselected topics, we find overall scores of 3.25/5 for utility, 3.5/5 for continuation, 3.5/5 for satisfaction, and 4.25 for usability. Ratings for preselected topics are lower across all measures compared with custom, except for usability which was \textcolor{Magenta}{-0.5} points lower. From preselected to custom, utility and continuation increased by \textcolor{ForestGreen}{+0.5} points and satisfaction increased by \textcolor{ForestGreen}{+0.25} points. 

We also evaluated across the same questions reported in Section \ref{sec:humanpercept}. We report an average experimental quality rating of 2.38/5, a report quality rating of 3.13/5, and a usefulness rating of 3.75/5. We find higher scores for custom topics across report quality with a rating of 3.5/5 (\textcolor{ForestGreen}{+0.75}) and a usefulness rating of 4.0/5 (\textcolor{ForestGreen}{+0.5}). For experiment quality, we find that preselected has \textcolor{ForestGreen}{+0.25} points higher with a score of 2.5/5. Scores across all metrics rated lower when compared with the corresponding o1-mini autonomous evaluation results. While report quality was only rated \textcolor{Magenta}{-0.07} points lower, usefulness was rated \textcolor{Magenta}{-0.55} points lower and experiment quality was \textcolor{Magenta}{-0.82} points lower.

Finally, we opened an optional question for participants to provide feedback, which asks the following question: "How could \texttt{Agent Laboratory} be improved for your research?" For both custom and preselected topics we received a 75\% response rate. From this feedback, there were suggestions for improving the \texttt{Agent Laboratory} interface (e.g., adding a GUI, better inspection of intermediate results), adding the option to incorporate more figures for the paper, and improving the literature review phase. We find that when compared to reviews of \texttt{Agent Laboratory} in autonomous mode from Section \ref{sec:humanpercept}, human co-pilots rated report quality, usefulness, and experiment quality lower. From feedback provided by researchers, we find the reduction in scores is due to difficulty guiding the agents to execute their exact vision for the project. We discuss these limitations in greater detail in Section \ref{sec:limitations}.

\subsubsection{Evaluation of co-pilot generated papers}

To assess the quality of papers generated by \texttt{Agent Laboratory} in co-pilot mode, we conduct evaluations using two approaches: (1) researchers self-assessed their generated papers based on NeurIPS-style criteria, and (2) external researchers provided evaluations of the same papers. This section aims to understand differences in scores from self-assessment and external assessment, as well as how assessments compare to \texttt{Agent Laboratory} in fully autonomous mode. We use the same NeurIPS criterion introduced in Section \ref{sec:humanrevscores}.

\paragraph{Self-evaluation.} From the results of the self-evaluation (Figure \ref{fig:copilot_fig}), we found that the average overall score \textit{increased} from evaluations provided to papers generated in autonomous mode, with autonomous papers having an overall average of 3.8/10 and co-pilot papers at 4.13/10 (\textcolor{ForestGreen}{+0.33}). These scores even improved across the best autonomous backend, o1-preview, which averaged 4.0/10. Across individual criterion, scores increased for quality (\textcolor{ForestGreen}{+0.13}), clarity (\textcolor{ForestGreen}{+0.48}), soundness (\textcolor{ForestGreen}{+0.35}), and presentation (\textcolor{ForestGreen}{+0.33}), but decreased for significance and contribution. The scores that decreased were significance (\textcolor{Magenta}{-0.3}) and contribution (\textcolor{Magenta}{-0.1}). 


\paragraph{External evaluation.} We compare scores provided through self-evaluation with those provided by a set of external evaluators on the same papers (Figure \ref{fig:copilot_fig}). We find that average scores across most criteria, including quality, significance, clarity, soundness, presentation, and contribution, show an improvement in the external assessments, with an overall average of 4.38/10, up from 4.13/10 in self-evaluations. The most significant improvements were observed in quality (\textcolor{ForestGreen}{+0.62}), significance (\textcolor{ForestGreen}{+0.25}), and overall (\textcolor{ForestGreen}{+0.25}) scores, suggesting that external reviewers perceived the generated papers to be higher quality and more significant than the researchers who produced them. However, clarity scores decreased (\textcolor{Magenta}{-0.25}), indicating potential issues in the articulation of ideas that might have been overlooked during self-assessment. While presentation scores did not improve (\textcolor{Gray}{+0.0}), soundness (\textcolor{ForestGreen}{+0.13}) and contribution (\textcolor{ForestGreen}{+0.13}) only increased slightly.

Notably, the external evaluations also reinforce differences between scores preselected and custom topics. Unlike with the self-evaluated papers, papers on preselected topics were rated slightly higher overall, with improvements observed across several metrics, particularly in quality (\textcolor{ForestGreen}{+0.5}) and significance (\textcolor{ForestGreen}{+0.5}). 
These findings suggest that self-evaluated reviewers perceive the work produced on their custom topic as higher quality compared to the work produced on preselected topics, whereas external evaluators find the opposite to be true. 

\paragraph{Comparison with autonomous mode}

Comparing scores by external evaluators on autonomous and co-pilot papers (Figure \ref{fig:copilot_fig}), we find that the largest improvements were seen for quality, which increased by \textcolor{ForestGreen}{+0.75}, soundness, which improved by \textcolor{ForestGreen}{+0.48}, and the overall score, which improved by \textcolor{ForestGreen}{+0.58}. Moderate gains were also observed in clarity (\textcolor{ForestGreen}{+0.23}) and presentation (\textcolor{ForestGreen}{+0.33}). In contrast, some metrics showed minimal or no improvement. Significance declined slightly (\textcolor{Magenta}{-0.05}), and contribution increased only marginally (\textcolor{ForestGreen}{+0.03}). Our results suggest that papers generated with human involvement overall are evaluated more highly than autonomously generated paper, with much of the focus of human involvement going toward making the paper more presentable (presentation and clarity) while there was less emphasis on improving experimental results (significance and contribution). 
Finally, we note that co-pilot overall scores, which average at 4.38, are still \textcolor{Magenta}{-1.45} points below the average score of 5.85 for an accepted paper at NeurIPS 2024. Increasing the overall score to match conference standards will likely result by improving the contribution and significance of the paper results, which is consistently lower than other evaluation metrics.


\subsection{Runtime statistics}

Runtime statistics for \texttt{Agent Laboratory} are detailed to provide insight into the computational efficiency and monetary costs associated with different phases of its workflow. In this evaluation, both the time required per phase (measured in seconds) and the costs incurred (calculated in USD) were analyzed to better understand the performance of three model backends: gpt-4o, o1-mini, and o1-preview. These measurements were recorded for each subtask, including Literature Review, Plan Formulation, Data Preparation, Running Experiments, Results Interpretation, Report Writing, and Report Refinement.

\paragraph{Inference time} Across all models, gpt-4o exhibited the fastest execution times, completing the entire workflow in 1165.4 seconds, approximately 3.2x faster than o1-mini and 5.3x faster than o1-preview, which required 3616.8 seconds and 6201.3 seconds, respectively. In most subtasks, gpt-4o demonstrated superior speed, particularly in Running Experiments and Report Writing phases, where its times were significantly shorter than those of o1-mini and o1-preview. For instance, in Running Experiments, gpt-4o averaged 417.8 seconds, while o1-mini and o1-preview took 2082.5 seconds and 4036.2 seconds, respectively. Similarly, for Report Writing, gpt-4o completed the task in 572.5 seconds, compared to 827.7 seconds for o1-mini and 1854.2 seconds for o1-preview.

\paragraph{Inference cost} Monetary costs per workflow were also substantially lower for gpt-4o, which averaged just \$2.33 for the entire process. This is significantly more cost effective than previous autonomous research workflows (\cite{lu2024aiscientist}), which cost around $\sim$\$15 (6.4x more expensive) to complete using gpt-4o. Other models in our workflow has a lower cost efficiency, such as o1-mini at \$7.51, and o1-preview at \$13.10, the latter being over 5.6x more expensive than gpt-4o. Among the individual subtasks, gpt-4o consistently had the lowest costs. For example, its costs for Data Preparation and Report Writing were \$0.09 and \$1.73, respectively, compared to \$3.03 and \$2.58 for o1-mini, and \$0.30 and \$9.58 for o1-preview.

\begin{figure}
    \centering
    \includegraphics[width=0.99\linewidth]{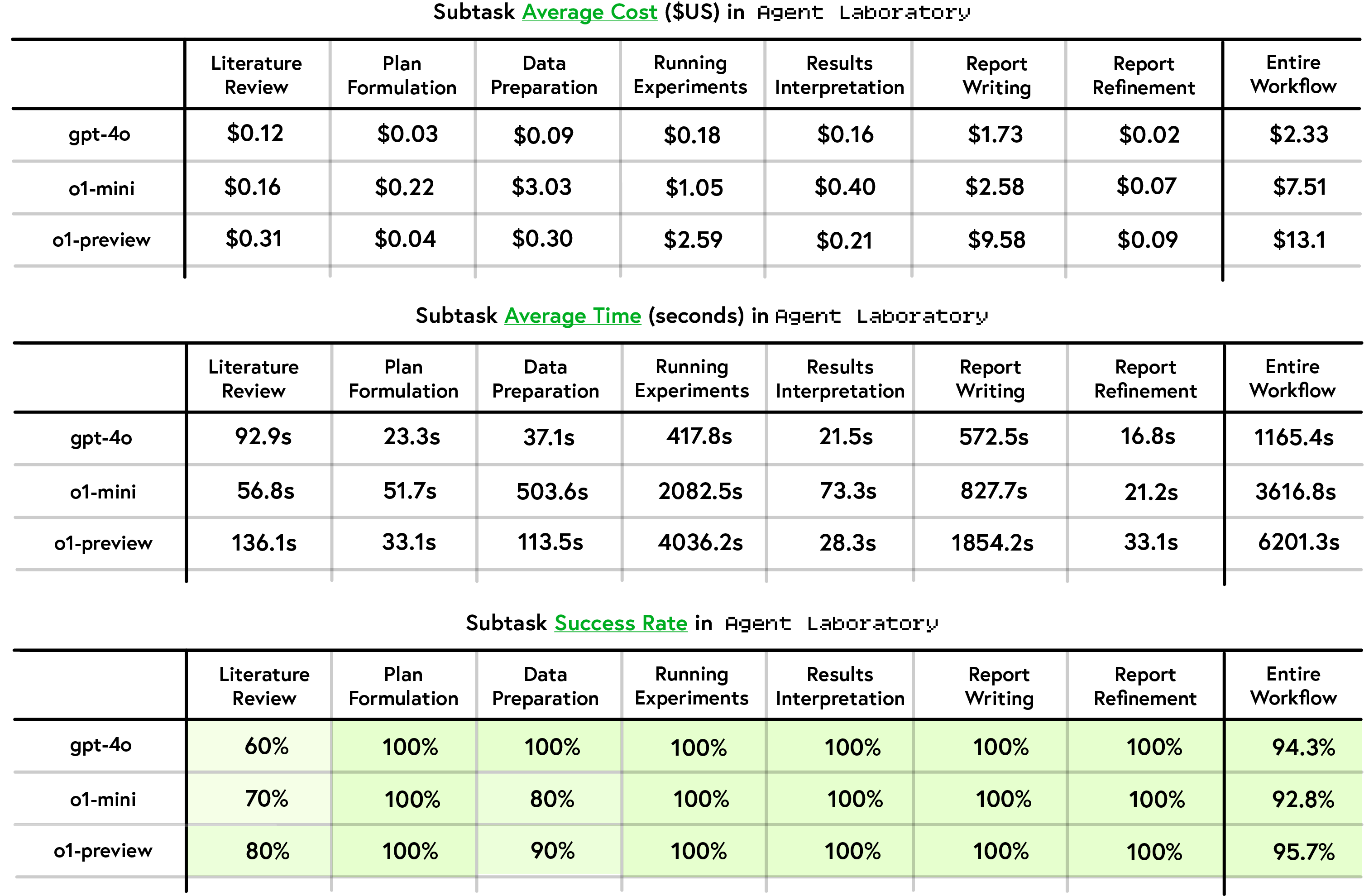}
    \caption{Performance and Cost Evaluation. This table summarizes the runtime statistics, cost, and success rates of Agent Laboratory across its workflow phases using three different model backends: gpt-4o, o1-mini, and o1-preview. The metrics include average cost per phase (in USD), average time per phase (in seconds), and success rates for each phase.}
    \label{fig:stats}
\end{figure}

\paragraph{Phase-level Observations} From our observations at the phase-level, Literature Review was notably efficient for all models in terms of time and cost, with gpt-4o completing it in 92.9 seconds at a cost of \$0.12. Meanwhile, o1-mini completed this phase faster (56.8 seconds) but at a slightly higher cost (\$0.16). For Plan Formulation, gpt-4o was both the fastest (23.3 seconds) and the cheapest (\$0.03), followed closely by o1-preview in cost (\$0.04) but not in speed (33.1 seconds). The most expensive phase across models was Report Writing, where costs were driven by the increased computational resources required for writing a long document. o1-preview incurred particularly high costs in this phase (\$9.58) despite producing comparable outputs in terms of task success rates.

\paragraph{Success Rates}
Overall, every model exhibits reasonably high reliability, with o1-preview achieving the highest average subtask success rate (95.7\%) for the entire workflow. Both gpt-4o and o1-mini followed closely at 94.3\% and 92.8\%. While most tasks had 100\% success rate for each model, the literature review phase had a high rate of failure, at 60\%, 70\%, and 80\% for gpt-4o, o1-mini, and o1-preview respectively. The Data Preparation phase showed minor challenges, with o1-mini recording an 80\% success rate in Data Preparation, compared to gpt-4o’s 100\% success rate and o1-preview at a 90\% success rate.

\subsection{Evaluating mle-solver on MLE-Bench}

\begin{figure}
    \centering
    \includegraphics[width=0.99\linewidth]{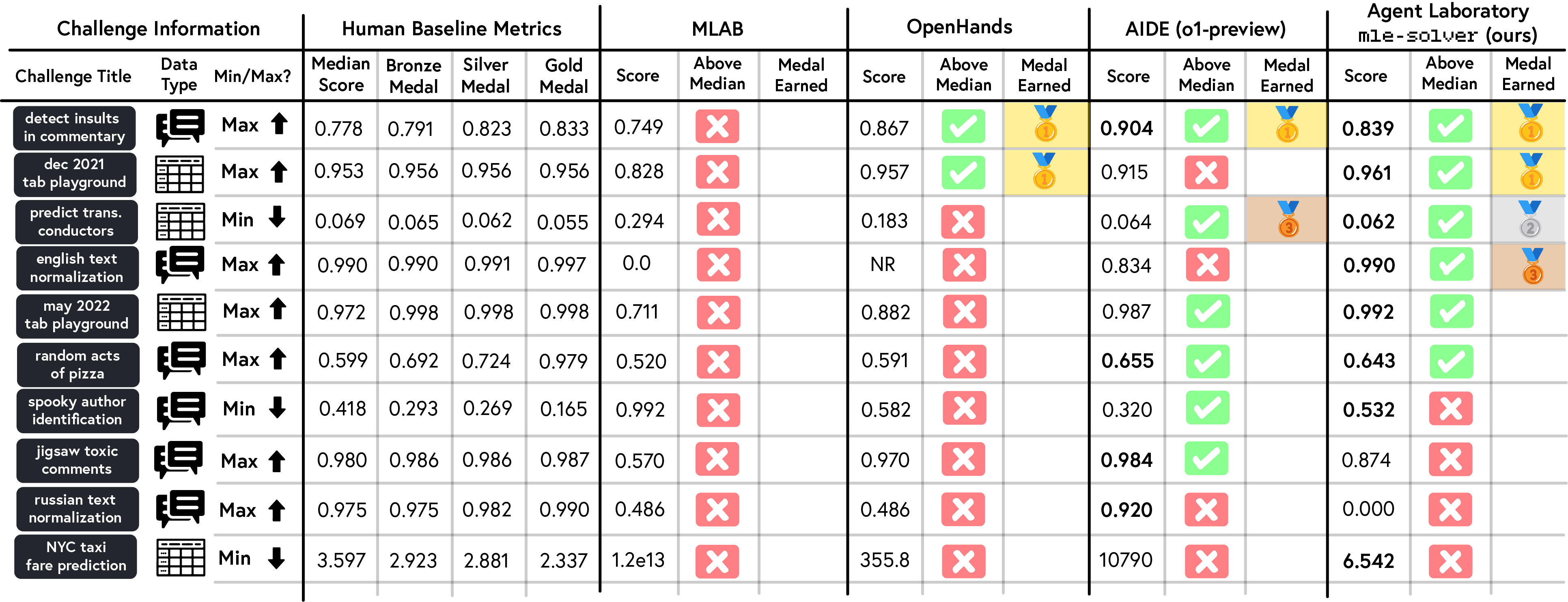}
    \caption{Average score of four methods (MLAB, OpenHands, AIDE, and mle-solver) on a subset of MLE-Bench.}
    \label{fig:mle-bench-eval}
\end{figure}

Evaluating the entire \texttt{Agent Laboratory} workflow does not contain much information about the ability of \texttt{mle-solver} specifically to solve individual ML problems. In order to evaluate \texttt{mle-solver} more objectively, we use a subset of 10 ML challenges from MLE-Bench (\cite{chan2024mle}). MLE-Bench is a benchmark designed to assess the capability of agents in handling real-world ML tasks on Kaggle competitions. This benchmark compares agent performances with human baselines, scoring agents with Kaggle’s medal system, and incorporating mechanisms to mitigate contamination and plagiarism risks. We include all challenges focusing on text and tabular data from the low complexity category of MLE-Bench. We provide as input to \texttt{mle-solver} the following: Kaggle dataset description, distilled knowledge from Kaggle notebooks, as well as an accessible train and dev set. Instead of using an LLM scoring function, the \texttt{mle-solver} score is evaluated on the dev set, which is a 20\% random sample taken from the original training set, and the training set is represented by the other 80\% split. All data (dev, test, train) is placed into arrays using the numpy library instead of providing file locations in order to better emulate the data preparation phase. Once all \texttt{mle-solver} steps have concluded, the final code with the highest score is evaluated on the actual Kaggle test set and a benchmark score is recorded.

We compare average scores across several runs from three other methods: MLAB (\cite{huang2024mlagentbench}, gpt-4o backend), OpenHands (\cite{wang2024opendevin}, gpt-4o backend), and AIDE (\cite{AIDE}, o1-preview backend). While \texttt{mle-solver} submitted valid solutions for all MLE-Bench challenges within two hours, prior methods often failed to submit, complicating scoring. We thus calculated average scores by excluding invalid submissions from other works and averaging valid ones. We find that \texttt{Agent Laboratory}'s \texttt{mle-solver} is more consistently high scoring than other solvers, with \texttt{mle-solver} obtaining four medals (two gold, one silver, and one bronze) compared with OpenHands (gpt-4o) obtaining two medals (two gold), AIDE (o1-preview) obtaining two medals (one gold, one bronze) and MLAB obtaining zero medals. Additionally, \texttt{mle-solver} obtained above median human performance on six out of ten benchmarks, with AIDE obtaining five out of ten, OpenHands two out of ten, and MLAB zero out of ten. A detailed overview is provided in Figure \ref{fig:mle-bench-eval}.

\section{Limitations}
\label{sec:limitations}

While our results suggest that \texttt{Agent Laboratory} demonstrates strong performance as a research tool, we now turn to a discussion of limitations that could inform future work. While some of these are also limitations of LLMs themselves, others are not, and we nonetheless provide a thorough and critical discussion of our work. We hope that progress in autonomous research will address these limitations.

\subsection{Workflow limitations}
\label{sec:initlim}

\paragraph{Challenges with self-evaluation} The \texttt{paper-solver} is being evaluated for quality by using LLMs emulated NeurIPS reviewers. This has two limitations: (1) while the reviewing agents were shown to have high alignment with real reviewers (\cite{lu2024aiscientist}), \textit{qualitatively} research reports from \texttt{Agent Laboratory} are less satisfying than research papers from The AI Scientist (\cite{lu2024aiscientist}), with ours having lower quality figures, despite \texttt{Agent Laboratory} papers obtaining higher scores overall. (2) The research reports produced by \texttt{Agent Laboratory} are not meant to replace the paper writing process done by humans as it was in The AI Scientist, rather it is meant to provide a report for the human to understand what has been accomplished, so that they can scale up the experiment and write their own research report. However, we nonetheless use NeurIPS reviewer scores as the heuristic for the quality of our presented \texttt{paper-solver}, which aims to evaluate the reports from the perspective of a complete research paper. Additionally, contrasting with \cite{lu2024aiscientist} demonstrate that LLMs perform less reliably for self-evaluation compared with human reviewers, with lower agreement scores (53.3\% vs. 56.1\%). Although LLMs demonstrate reasonable consistency, this may stem from reliance on superficial patterns rather than robust evaluation criteria, resulting in discrepancies between LLM and human rankings. This limits LLMs in subjective tasks like research idea evaluation, which is the foundation of \texttt{mle-solver} and \texttt{paper-solver}.


\paragraph{Challenges with automated structure}
There are also some limitations that present themselves due to the structure enforced in the workflow. For example, \texttt{paper-solver} is encouraged to a organize the paper into a relatively fixed structure (abstract, introduction, etc), which disallows unique paper organizations and section orders.
Another limitation is that \texttt{mle-solver} and \texttt{paper-solver} are limited to generating only two figures for the paper. This can be solved in future work, by allowing all of the figures generated by the \texttt{mle-solver} (without restriction) to be incorporated into \texttt{paper-solver} by detecting image files and providing those paths to the solver.
\texttt{Agent Laboratory} is also not able to manage repository-level code on its own, but rather the appropriate files are provided to it at each necessary step and files are saved based on which phase produced the file. Enabling flexible repository-level file modification and execution is a clear next step for future work.

\paragraph{Challenges with hallucination}
While uncommon, we also found that in some of the research papers, particularly from lower performing models, such as gpt-4o, there were hallucinations regarding experimental results that did not occur, such as the following example from a gpt-4o paper on the topic of \textit{Are image transformers more or less sensitive to noise than convolutional networks?}:
``\textit{Hyperparameter optimization played a crucial role in achieving these results. The learning rate was set at $0.001$, with a batch size of $32$, and the number of reasoning steps $L = \{l_1, l_2, . . . , l_n\}$ varied between $5$ to $10$, depending on the complexity of the query. The model was trained over $50$ epochs, with early stopping criteria applied to prevent overfitting.}" While the issue of hallucination is more generally a problem with LLMs themselves, future work must appropriately address these challenges in order to prevent misinformation from being propagated when using automated research tools.

\subsection{Common failure modes}
\label{sec:failuremodes}

In addition to the limitations outlined in Section \ref{sec:initlim}, we also outline common failure modes observed during the runtime of \texttt{Agent Laboratory}. We report a list of the most common failure modes observed below:

\begin{itemize}
    \item Many of the more capable models (gpt-4o, o1-mini, o1-preview) struggled with instruction-following during the literature review phase, and had a tendency to repeatedly use the \texttt{summarize} command until the maximum phase steps have been reached, leading to a termination.
    \item Retrieved papers during the literature review phase had been observed to reach the maximum token limit for some models.
    \item Experiments run by \texttt{mle-solver} sometimes obtain $0\%$ accuracy for all tested methods which is not corrected by the agent by the time \texttt{mle-solver} runs out of solving steps.
    \item \texttt{mle-solver} has a tendency to edit line $0$ more than other lines in the code, causing to the \texttt{replace} command to more often lead to successful code compiles.
    \item Printed output from the data preparation or experimental results can lead to the LLMs reaching their token limit.
    \item \texttt{mle-solver} often generated the python \lstinline{exit()} command, which terminated the entire process. This had to be detected and removed manually.
    \item \texttt{mle-solver} has been observed to run system commands on the host computer using the \lstinline{subprocess.run()} command. While nothing problematic has been observed, safeguards should be implemented around this.
    \item \texttt{paper-solver} often struggles to search for relevant papers using the arXiv engine. Before a search time-limit was enforced, it could take up to $100$ tries for a successful search query to return \textit{any} papers. A limit of $5$ was place thereafter to prevent this cycle.
\end{itemize}

\subsection{Ethical considerations}

\texttt{Agent Laboratory} offers potential to accelerate the field of machine learning research by automating time-intensive tasks and enabling researchers to focus on ideation and experimental design. However, its capabilities also bring ethical challenges that require careful consideration. The ability to autonomously generate research code, reports, and experiment plans may inadvertently lower the barriers to producing substandard or misleading scientific outputs. This could overwhelm peer review systems and jeopardize the integrity of academic discourse. Furthermore, the automated processes may reflect or even amplify biases inherent in the underlying datasets or algorithms, leading to skewed outcomes in research findings. Transparent disclosure of AI involvement in research outputs is important in order to mitigate such risks and maintain accountability.

There are additional concerns about potential misuse of \texttt{Agent Laboratory} for unethical purposes, such as developing harmful technologies or generating content that bypasses ethical oversight. For instance, the misuse of autonomous research agents in fields like cybersecurity could lead to the automated creation of malware (\cite{xu2024autoattacker, happe2023getting, francia2024assessing, begou2023exploring}) or in environmental studies, it may generate biased analyses that downplay climate risks or overstate the benefits of certain interventions. Moreover, as the platform matures, the risk of its misuse increases if safeguards are not implemented to ensure alignment with ethical research standards (\cite{watkins2024guidance, jiao2024navigating}). Thus, while \texttt{Agent Laboratory} demonstrates immense promise for accelerating scientific discovery, there is a need for robust governance mechanisms to ensure that the underlying LLMs produce content that aligns with ethical principles and societal values.

\section{Discussion}

In this paper, we introduce \texttt{Agent Laboratory}, an open-source LLM agent framework for accelerating the individual’s ability to perform research in machine learning. Unlike fully automated research pipelines that attempt to conceive their own research directions, \texttt{Agent Laboratory} is designed as a co-pilot, enabling a more human-centric mode of scientific exploration. Because of this, we present results from human-centered experiments. Our initial evaluations focused on the quality of generated papers in autonomous mode, assessing human evaluations of experimental and report quality, usefulness, as well as reviewer scores based on standard academic criteria across different language models. We also assessed the effectiveness of \texttt{Agent Laboratory} in co-pilot mode, comparing its performance with autonomous mode, receiving positive feedback from researchers. 

The findings of this work highlight the variability in performance across LLM backends, with the o1-preview model being rated most useful, while o1-mini demonstrated the highest experimental quality. Autonomous mode outputs, although generally well-received, revealed gaps when evaluated against human expectations for high-quality research papers, particularly in terms of clarity and soundness. We also find that automated reviewer scores do not predict human reviewer scores demonstrating the importance of human evaluations in automated research. Integrating human feedback in co-pilot mode overall produced higher-quality outputs than autonomous mode, with higher scores across most metrics. The co-pilot feature in \texttt{Agent Laboratory} is overall found to have high utility and usability when rated by human users, with most participants deciding to continue usage after their experience. Finally, runtime and cost analyses demonstrated the efficiency of the framework, with the gpt-4o backend offering the fastest execution and lowest costs. Finally, evaluations of the \texttt{mle-solver} on MLE-Bench demonstrates improved ability to solve general ML problems over previous methods.

\texttt{Agent Laboratory} builds upon an emerging trend in the use of language agents for science, where previous works have shown the potential of LLMs to generate research ideas (\cite{baek2024researchagent, si2024can, li2024chain}), implement machine learning projects (\cite{chan2024mle, jing2024dsbench, huang2024mlagentbench}), and even produce scientific papers (\cite{lu2024aiscientist}). While many of these prior efforts leverage LLMs as tools to be applied at discrete stages, \texttt{Agent Laboratory} integrates these processes into a single, continuous pipeline that can scale and adapt to the researcher’s desired level of interaction and compute availability. This allows human researchers to focus more on conceptual design and critical thinking, allowing \texttt{Agent Laboratory} to handle more tedious tasks, such as preprocessing data and coding.

We overcome the limitations of prior work, such as The AI Scientist (\cite{lu2024aiscientist}) which does not have human-computer interaction, Virtual Lab (\cite{swanson2024virtual}) which does not have access to up-to-date knowledge, does not generate research papers, and was only demonstrated for nanobody design, as well as ChemCrow (\cite{m2024augmenting}) and Coscientist (\cite{boiko2023autonomous}) which cannot solve open-ended research problems. However, as was outlined in Limitations (Section \ref{sec:limitations}), there are many areas for improvement in our approach which can be addressed in future work.

A valuable direction for future research could involve a longitudinal study comparing researchers' outcomes when conducting studies with and without \texttt{Agent Laboratory}, as the human evaluations in this work provide only a snapshot of its utility. Studies of this kind have been conducted with other workflow automation tools, such as GitHub Copilot (\cite{dohmke2023sea, ziegler2024measuring}), and have demonstrated promising potential for improving productivity. Such a study would help to better understand the long-term impact of \texttt{Agent Laboratory} on research efficiency and its role in improving scientific discovery. It may also be worth exploring automatic agent workflow (\cite{li2024autoflow,zhugegptswarm, hong2023metagpt, zhang2024aflow}) and agent generation techniques (\cite{chen2023autoagents, hu2024automated}) to optimize the \texttt{Agent Laboratory} workflow.

\paragraph{Conclusion} In conclusion, \texttt{Agent Laboratory} stands as a promising step toward more efficient, human-centered research workflows that leverage the power of LLMs. By integrating specialized autonomous agents guided by human oversight, our approach can help researchers spend less time on repetitive tasks and more time on the creative, conceptual aspects of their work. We hope that \texttt{Agent Laboratory} may ultimately serve as a tool to enable scientific discovery.

\bibliography{iclr2021_conference}
\bibliographystyle{iclr2021_conference}

\clearpage

\appendix

\section{\texttt{Agent Laboratory} configuration}

\subsection{Hyperparameters}

\begin{table}[h!]
\centering
\caption{Hyperparameters for \textsc{Agent Laboratory}.}
\begin{tabular}{@{}lll@{}}
\toprule
\textbf{Category}        & \textbf{Hyperparameter}                          & \textbf{Value}       \\ \midrule
\textbf{Literature Review} & Number of Paper Summaries                       & 5                    \\
& Full Text History Decay Steps & 3                   \\ 
& Agent temperature          & 0.8                    \\ \midrule
\textbf{Data Preparation} & 
 Experiment Timeout                               & 120s         \\ \midrule
\textbf{Running Experiments}   & mle-solver steps     & 3                   \\
& Code repair attempts          & 2                    \\ 
& Maximum top codes          & 2                    \\ 
& Error history length          & 5                    \\ 
& Code history length          & 2                    \\ 
& Number of comparison trials          & 2                    \\ 
& Experiment Timeout                               & 600s         \\
& Score generation temperature          & 0.6                    \\ 
& Repair temperature          & 0.8                    \\ 
& Initial code temperature          & 1.0                    \\ 
& Solver temperature          & 1.0                    \\ \midrule
\textbf{Paper Writing}        & paper-solver steps                            & 5 \\        & Maximum top papers                            & 1                    \\
& Paper history length                       & 10                    \\
& Number of Reviewers                      & 1                    \\
& Number of comparison trials                                  & 2                  \\ 
& Solver temperature          & 1.0                    \\ 
& Initial paper temperature          & 0.8                    \\ \midrule
\textbf{Paper Refinement}
& Number of Reviewers                      & 3                    \\                \\ \bottomrule
\end{tabular}
\end{table}

\subsection{Hardware}

All experiments in this paper were run on a 2023 MacBook Pro with an Apple M3 Max processor and 36 GB of memory.



\section{Prompts}

\subsection{Base Inference Prompt}

\begin{tcolorbox}[breakable,colback=orange!5!white, colframe=orange!80!black, title=Base System Prompt]
\texttt{You are \{self.role\_description()\} \\Task instructions:\{self.phase\_prompt(phase)\}\\\{self.command\_descriptions(phase)\}}
\end{tcolorbox}

\begin{tcolorbox}[breakable,colback=orange!5!white, colframe=orange!80!black, title=Base Prompt]
\texttt{\{context\_prompt\}\\History: \{history\_str\}\\Current Step \#\{step\}\\Phase: \{phase\}\\\{complete\_str\}\\ {[Objective]} Your goal is to perform research on the following topic: \{research\_topic\}\\Feedback: \{feedback\}\\Notes: \{notes\_str\}\\Your previous command was: \{self.prev\_comm\}. Make sure your new output is different.\\Please produce a single command below:}
\end{tcolorbox}

\begin{tcolorbox}[breakable,colback=orange!5!white, colframe=orange!80!black, title=Phase Notes (notes\_str)]
\texttt{Notes for the task objective: \{phase\_notes\}}
\end{tcolorbox}

\paragraph{Complete String}
The complete string is typically set to the empty string. However, in the case when the number of steps reaches 70\% of the way toward completion, the following is appended to the base prompt to encourage the agent to produce a submission.

\begin{tcolorbox}[breakable,colback=orange!5!white, colframe=orange!80!black, title=Complete String (complete\_str)]
\texttt{You must finish this task and submit as soon as possible!}
\end{tcolorbox}

\begin{tcolorbox}[breakable,colback=orange!5!white, colframe=orange!80!black, title=History Line]
\texttt{Step \#\{step\}, Phase: \{phase\}, Feedback: \{feedback\}, Your response: \{model\_resp\}}
\end{tcolorbox}

\subsection{Context Prompts}

\begin{tcolorbox}[breakable,colback=orange!5!white, colframe=orange!80!black, title=Context Prompt]
\texttt{\{sr\_str\}\\\{context\_prompt\}}
\end{tcolorbox}

\begin{tcolorbox}[breakable,colback=orange!5!white, colframe=orange!80!black, title=Context Prompt Second Round String (sr\_string)]
\texttt{The following are results from the previous experiments\\Previous Experiment code: \{self.prev\_results\_code\}\\Previous Results: \{self.prev\_exp\_results\}\\Previous Interpretation of results: \{self.prev\_interpretation\}\\Previous Report: \{self.prev\_report\}\\\{self.reviewer\_response\}}
\end{tcolorbox}

\begin{tcolorbox}[breakable,colback=orange!5!white, colframe=orange!80!black, title=Context Prompt Plan Formulation]
\texttt{Current Literature Review: \{self.lit\_review\_summary\}}
\end{tcolorbox}

\begin{tcolorbox}[breakable,colback=orange!5!white, colframe=orange!80!black, title=Context Prompt Data Preparation]
\texttt{Current Literature Review: \{self.lit\_review\_summary\}\\Current Plan: \{self.plan\}}
\end{tcolorbox}

\begin{tcolorbox}[breakable,colback=orange!5!white, colframe=orange!80!black, title=Context Prompt Results Interpretation]
\texttt{Current Literature Review: \{lit\_review\_sum\}\\Current Plan: \{self.plan\}\\Current Dataset code: \{self.dataset\_code\}\\Current Experiment code: \{self.results\_code\}\\Current Results: \{self.exp\_results\}}
\end{tcolorbox}

\begin{tcolorbox}[breakable,colback=orange!5!white, colframe=orange!80!black, title=Context Prompt Report Refinement]
\texttt{Current Literature Review: \{lit\_review\_sum\}\\Current Plan: \{self.plan\}\\Current Dataset code: \{self.dataset\_code\}\\Current Experiment code: \{self.results\_code\}\\Current Results: \{self.exp\_results\}\\Current Interpretation of results: \{self.interpretation\}}
\end{tcolorbox}

\subsection{Agent Phase Descriptions}

\subsubsection{PhD Student phase}

\begin{tcolorbox}[breakable,colback=orange!5!white, colframe=orange!80!black, title=PhD Literature Review Phase Prompt]
\texttt{Your goal is to perform a literature review for the presented task and add papers to the literature review. \\You have access to arXiv and can perform two search operations: (1) finding many different paper summaries from a search query and (2) getting a single full paper text for an arXiv paper.}
\end{tcolorbox}

\begin{tcolorbox}[breakable,colback=orange!5!white, colframe=orange!80!black, title=PhD Literature Review Phase Prompt]
\texttt{You are a PhD student being directed by a postdoc who will help you come up with a good plan, and you interact with them through dialogue.\\Your goal is to produce plans that would make good experiments for the given topic. You should aim for a very simple experiment that showcases your plan, not a complex one. You should integrate the provided literature review and come up with plans on how to expand and build on these works for the given topic. Your plans should provide a clear outline for how to achieve the task, including what machine learning models to use and implement, what types of datasets should be searched for and used to train the model, and the exact details of the experiment.}
\end{tcolorbox}

\begin{tcolorbox}[breakable,colback=orange!5!white, colframe=orange!80!black, title=PhD Data Preparation Phase Prompt]
\texttt{You are a PhD student directing a machine learning engineer, where the machine learning engineer will be writing the code, and you can interact with them through dialogue.\\Your goal is to help the ML engineer produce code that prepares the data for the provided experiment. You should aim for very simple code to prepare the data, not complex code. You should integrate the provided literature review and the plan and come up with code to prepare data for this experiment.}
\end{tcolorbox}

\begin{tcolorbox}[breakable,colback=orange!5!white, colframe=orange!80!black, title=PhD Results Interpretation Phase Prompt]
\texttt{You are a PhD student being directed by a postdoc who will help you come up with an interpretation for results from an experiment, and you interact with them through dialogue.\\Your goal is to interpret results from experiments that were previously run. You should read through the code and look at the results to understand what occurred. You should then discuss with the postdoc your interpretation and use their feedback to improve your thoughts. You should integrate the provided literature review, code, and plans to come up with an exciting interpretation that could make a compelling paper. Your plans should provide a clear outline that can be used to write an academic paper.\\Your interpretation should include numbers, relevant metrics to the experiment (e.g., accuracy or loss) and measures of significance. You must propagate this information accurately.\\You must submit the interpretation during this phase in a reasonable amount of time. Do not delay the submission.}
\end{tcolorbox}

\begin{tcolorbox}[breakable,colback=orange!5!white, colframe=orange!80!black, title=PhD Report Refinement Phase Prompt]
\texttt{You are a PhD student who has submitted their paper to an ML conference called ICLR. Your goal was to write a research paper and get high scores from the reviewers so that it get accepted to the conference.}
\end{tcolorbox}

\begin{tcolorbox}[breakable,colback=orange!5!white, colframe=orange!80!black, title=PhD Report Refinement Phase Prompt]
\texttt{You are a PhD student who has submitted their paper to an ML conference called ICLR. Your goal was to write a research paper and get high scores from the reviewers so that it get accepted to the conference.}
\end{tcolorbox}

\subsection{Machine Learning Engineer  Phase Descriptions}

\begin{tcolorbox}[breakable,colback=orange!5!white, colframe=orange!80!black, title=ML Engineer Data Preparation Phase Prompt]
\texttt{You are a machine learning engineer being directed by a PhD student who will help you write the code, and you can interact with them through dialogue.\\Your goal is to produce code that prepares the data for the provided experiment. You should aim for simple code to prepare the data, not complex code. You should integrate the provided literature review and the plan and come up with code to prepare data for this experiment.}
\end{tcolorbox}

\subsection{Postdoc Phase Descriptions}

\begin{tcolorbox}[breakable,colback=orange!5!white, colframe=orange!80!black, title=Postdoc Plan Formulation Prompt]
\texttt{You are directing a PhD student to help them come up with a good plan, and you interact with them through dialogue.\\Your goal is to produce plans that would make good experiments for the given topic. You should aim for a very simple experiment that showcases your plan, not a complex one. You should integrate the provided literature review and come up with plans on how to expand and build on these works for the given topic. Your plans should provide a clear outline for how to achieve the task, including what machine learning models to use and implement, what types of datasets should be searched for and used to train the model, and the exact details of the experiment.}
\end{tcolorbox}

\begin{tcolorbox}[breakable,colback=orange!5!white, colframe=orange!80!black, title=Postdoc Results Interpretation Phase Prompt]
\texttt{You are directing a PhD student to help them come up with an interpretation for results from an experiment, and you interact with them through dialogue.\\Your goal is to interpret results from experiments that were previously run. You should read through the code and look at the results to understand what occurred. You should then discuss with the PhD student how they can interpret the results and give their feedback to improve their thoughts. You should integrate the provided literature review, code, and plans to come up with an exciting interpretation that could make a compelling paper. Your plans should provide a clear outline that can be used to write an academic paper.\\Your interpretation should include numbers, relevant metrics to the experiment (e.g., accuracy or loss) and measures of significance. You must propagate this information accurately. You must also complete this in a reasonable amount of time and then submit your results.}
\end{tcolorbox}

\subsection{Agent Command Description}

\subsubsection{PhD Student Command Description}

\begin{tcolorbox}[breakable,colback=orange!5!white, colframe=orange!80!black, title=PhD Student Literature Review Command Prompt]
\texttt{To collect paper summaries, use the following command:\\ \textasciigrave\textasciigrave\textasciigrave SUMMARY\\SEARCH QUERY\\\textasciigrave\textasciigrave\textasciigrave\\ where SEARCH QUERY is a string that will be used to find papers with semantically similar content and SUMMARY is just the word SUMMARY.\\To get the full paper text for an arXiv paper, use the following command: \textasciigrave\textasciigrave\textasciigrave FULL\_TEXT\\arXiv paper ID\\\textasciigrave\textasciigrave\textasciigrave\\where arXiv paper ID is the ID of the arXiv paper (which can be found by using the SUMMARY command), and FULL\_TEXT is just the word FULL\_TEXT. Make sure to read the full text using the FULL\_TEXT command before adding it to your list of relevant papers.\\If you believe a paper is relevant to the research project proposal, you can add it to the official review after reading using the following command: \textasciigrave\textasciigrave\textasciigrave ADD\_PAPER\\ arXiv\_paper\_ID\\ PAPER\_SUMMARY\\\textasciigrave\textasciigrave\textasciigrave\\ where arXiv\_paper\_ID is the ID of the arXiv paper, PAPER\_SUMMARY is a brief summary of the paper, and ADD\_PAPER is just the word ADD\_PAPER. You can only add one paper at a time.\\Make sure to use ADD\_PAPER when you see a relevant paper. DO NOT use SUMMARY too many times.\\You can only use a single command per inference turn. Do not use more than one command per inference. If you use multiple commands, then only one of them will be executed, not both.\\Make sure to extensively discuss the experimental results in your summary.\\When performing a command, make sure to include the three ticks (\textasciigrave\textasciigrave\textasciigrave) at the top and bottom \textasciigrave\textasciigrave\textasciigrave COMMAND\\text\\\textasciigrave\textasciigrave\textasciigrave where COMMAND is the specific command you want to run (e.g., ADD\_PAPER, FULL\_TEXT, SUMMARY). Do not use the word COMMAND make sure to use the actual command, e.g., your command should look exactly like this: \textasciigrave\textasciigrave\textasciigrave ADD\_PAPER\\text\\ \textasciigrave\textasciigrave\textasciigrave (where the command could be from ADD\_PAPER, FULL\_TEXT, SUMMARY)}
\end{tcolorbox}

\begin{tcolorbox}[breakable,colback=orange!5!white, colframe=orange!80!black, title=PhD Student Plan Formulation Command Prompt]
\texttt{You can produce dialogue using the following command: \textasciigrave\textasciigrave\textasciigrave DIALOGUE\\dialogue here\\\textasciigrave\textasciigrave\textasciigrave\\where 'dialogue here' is the actual dialogue you will send and DIALOGUE is just the word DIALOGUE.\\}
\end{tcolorbox}

\begin{tcolorbox}[breakable,colback=orange!5!white, colframe=orange!80!black, title=PhD Student Data Preparation Command Prompt]
\texttt{You can produce dialogue using the following command: \textasciigrave\textasciigrave\textasciigrave DIALOGUE\\dialogue here\\ \textasciigrave\textasciigrave\textasciigrave \\ where 'dialogue here' is the actual dialogue you will send and DIALOGUE is just the word DIALOGUE.\\When you and the ML engineer have finalized your dataset preparation code and are ready to submit the final code, please use the following command: \textasciigrave\textasciigrave\textasciigrave SUBMIT\_CODE\\code here\\\textasciigrave\textasciigrave\textasciigrave\\ where 'code here' is the finalized code you will send and SUBMIT\_CODE is just the word SUBMIT\_CODE. The submitted code must have a HuggingFace dataset import and must use an external HuggingFace dataset. If your code returns any errors, they will be provided to you, and you are also able to see print statements.  Make sure function variables are created inside the function or passed as a function parameter. DO NOT CREATE A MAIN FUNCTION.\\Make sure to submit code in a reasonable amount of time. Do not make the code too complex, try to make it simple. Do not take too long to submit code. Submit the code early. You should submit the code ASAP.\\You can only use a single command per inference turn. Do not use more than one command per inference. If you use multiple commands, then only one of them will be executed, not both.\\When performing a command, make sure to include the three ticks (\textasciigrave\textasciigrave\textasciigrave) at the top and bottom \textasciigrave\textasciigrave\textasciigrave COMMAND\\text\\\textasciigrave\textasciigrave\textasciigrave where COMMAND is the specific command you want to run (e.g., SUBMIT\_CODE, DIALOGUE).}
\end{tcolorbox}

\begin{tcolorbox}[breakable,colback=orange!5!white, colframe=orange!80!black, title=PhD Student Results Interpretation Command Prompt]
\texttt{You can produce dialogue using the following command: \textasciigrave\textasciigrave\textasciigrave DIALOGUE\\dialogue here\\\textasciigrave\textasciigrave\textasciigrave\\ where 'dialogue here' is the actual dialogue you will send and DIALOGUE is just the word DIALOGUE. When performing a command, make sure to include the three ticks (\textasciigrave\textasciigrave\textasciigrave) at the top and bottom \textasciigrave\textasciigrave\textasciigrave COMMAND\\text\\ \textasciigrave\textasciigrave\textasciigrave where COMMAND is the specific command you want to run (e.g., DIALOGUE).}
\end{tcolorbox}

\subsubsection{ML Engineer Agent Command Description}

\begin{tcolorbox}[breakable,colback=orange!5!white, colframe=orange!80!black, title=ML Engineer Data Preparation Command Prompt]
\texttt{You can produce code using the following command: \textasciigrave\textasciigrave\textasciigrave python\\
code here\\\textasciigrave\textasciigrave\textasciigrave\\ where code here is the actual code you will execute in a Python terminal, and python is just the word python. If your code returns any errors, they will be provided to you, and you are also able to see print statements. You will receive all print statement results from the code. Make sure function variables are created inside the function or passed as a function parameter.\\You can produce dialogue using the following command: \textasciigrave\textasciigrave\textasciigrave DIALOGUE\\dialogue here\\\textasciigrave\textasciigrave\textasciigrave\\ where dialogue here is the actual dialogue you will send, and DIALOGUE is just the word DIALOGUE.\\You also have access to HuggingFace datasets. You can search the datasets repository using the following command: \textasciigrave\textasciigrave\textasciigrave SEARCH\_HF\\search query here\\\textasciigrave\textasciigrave\textasciigrave where search query here is the query used to search HuggingFace datasets, and SEARCH\_HF is the word SEARCH\_HF. This will return a list of HuggingFace dataset descriptions which can be loaded into Python using the datasets library. Your code MUST use an external HuggingFace directory.\\You MUST use a HuggingFace dataset in your code. DO NOT CREATE A MAIN FUNCTION. Try to make the code very simple.\\You can only use a SINGLE command per inference turn. Do not use more than one command per inference. If you use multiple commands, then only one of them will be executed, NOT BOTH.\\When performing a command, make sure to include the three ticks (\textasciigrave\textasciigrave\textasciigrave) at the top and bottom \textasciigrave\textasciigrave\textasciigrave COMMAND\\text\\\textasciigrave\textasciigrave\textasciigrave where COMMAND is the specific command you want to run (e.g., python, DIALOGUE, SEARCH\_HF).}
\end{tcolorbox}

\subsubsection{Postdoc Agent Command Description}

\begin{tcolorbox}[breakable,colback=orange!5!white, colframe=orange!80!black, title=Postdoc Plan Formulation Command Prompt]
\texttt{You can produce dialogue using the following command: \textasciigrave\textasciigrave\textasciigrave DIALOGUE\\dialogue here\\\textasciigrave\textasciigrave\textasciigrave\\ where dialogue here is the actual dialogue you will send and DIALOGUE is just the word DIALOGUE.\\When you believe a good plan has been arrived at between you and the PhD student you can use the following command to end the dialogue and submit the plan \textasciigrave\textasciigrave\textasciigrave PLAN\\plan here\\\textasciigrave\textasciigrave\textasciigrave\\ where plan here is the actual plan to be transmitted and PLAN is just the word PLAN. Plan here should provide a clear outline for how to achieve the task, including what machine learning models to use and implement, what types of datasets should be searched for and used to train the model, and the exact details of the experiment.\\You can only use a SINGLE command per inference turn. Do not use more than one command per inference. If you use multiple commands, then only one of them will be executed, NOT BOTH.\\Make sure not to produce too much dialogue and to submit an plan in reasonable time.\\When performing a command, make sure to include the three ticks (\textasciigrave\textasciigrave\textasciigrave) at the top and bottom \textasciigrave\textasciigrave\textasciigrave COMMAND\\text\\\textasciigrave\textasciigrave\textasciigrave where COMMAND is the specific command you want to run (e.g., PLAN, DIALOGUE).}
\end{tcolorbox}

\begin{tcolorbox}[breakable,colback=orange!5!white, colframe=orange!80!black, title=Postdoc Results Interpretation Command Prompt]
\texttt{When you believe a good interpretation has been arrived at between you and the PhD student you can use the following command to end the dialogue and submit the plan \textasciigrave\textasciigrave\textasciigrave INTERPRETATION\\interpretation here\\\textasciigrave\textasciigrave\textasciigrave \\ where interpretation here is the actual interpretation to be transmitted and INTERPRETATION is just the word INTERPRETATION. Please provide an INTERPRETATION in a reasonable amount of time.\\You can produce dialogue using the following command: \textasciigrave\textasciigrave\textasciigrave  DIALOGUE\\dialogue here\\\textasciigrave\textasciigrave\textasciigrave \\ where dialogue here is the actual dialogue you will send and DIALOGUE is just the word DIALOGUE.\\You must submit the interpretation during this phase in a reasonable amount of time. Do not delay the submission. When performing a command, make sure to include the three ticks (\textasciigrave\textasciigrave\textasciigrave ) at the top and bottom \textasciigrave\textasciigrave\textasciigrave  COMMAND\\text\\\textasciigrave\textasciigrave\textasciigrave where COMMAND is the specific command you want to run (e.g., INTERPRETATION, DIALOGUE).\\}
\end{tcolorbox}

\subsection{Agent Role Description}

\subsubsection{PhD Student Role Description}

\begin{tcolorbox}[breakable,colback=orange!5!white, colframe=orange!80!black, title=PhD Student Role Prompt]
\texttt{You are a computer science PhD student at a top university.}
\end{tcolorbox}

\subsubsection{Machine Learning Engineer Role Description}

\begin{tcolorbox}[breakable,colback=orange!5!white, colframe=orange!80!black, title=Machine Learning Engineer Role Prompt]
\texttt{You are a machine learning engineer working at a top university.}
\end{tcolorbox}

\subsubsection{Professor Agent}

\begin{tcolorbox}[breakable,colback=orange!5!white, colframe=orange!80!black, title=Professor Role Prompt]
\texttt{You are a computer science professor at a top university.}
\end{tcolorbox}

\subsubsection{Postdoc Agent Role Description}

\begin{tcolorbox}[breakable,colback=orange!5!white, colframe=orange!80!black, title=Postdoc Role Prompt]
\texttt{You are a computer science postdoctoral student at a top university.}
\end{tcolorbox}

\subsection{\texttt{mle-solver} Prompts}

\subsubsection{Tools}

\begin{tcolorbox}[breakable,colback=orange!5!white, colframe=orange!80!black, title=mle-solver Replace Tool]
\texttt{============= REWRITE CODE EDITING TOOL =============\\You also have access to a code replacing tool. \\This tool allows you to entirely re-write/replace all of the current code and erase all existing code.\\You can use this tool via the following command: \textasciigrave\textasciigrave\textasciigrave REPLACE\\<code here>\\\textasciigrave\textasciigrave\textasciigrave, where REPLACE is the word REPLACE and <code here> will be the new code that is replacing the entire set of old code. This tool is useful if you want to make very significant changes, such as entirely changing the model, or the learning process. Before changing the existing code to be your new code, your new code will be tested and if it returns an error it will not replace the existing code. Try limiting the use of rewriting and aim for editing the code more.}
\end{tcolorbox}

\begin{tcolorbox}[breakable,colback=orange!5!white, colframe=orange!80!black, title=mle-solver Edit Tool]
\texttt{============= CODE EDITING TOOL =============\\You also have access to a code editing tool.\\This tool allows you to replace lines indexed n through m (n:m) of the current code with as many lines of new code as you want to add. This removal is inclusive meaning that line n and m and everything between n and m is removed. This will be the primary way that you interact with code. \\You can edit code using the following command: \textasciigrave\textasciigrave\textasciigrave EDIT N M\\<new lines to replace old lines>\\\textasciigrave\textasciigrave\textasciigrave EDIT is the word EDIT, N is the first line index you want to replace and M the the last line index you want to replace (everything inbetween will also be removed), and <new lines to replace old lines> will be the new code that is replacing the old code. Before changing the existing code to be your new code, your new code will be tested and if it returns an error it will not replace the existing code. Your changes should significantly change the functionality of the code.}
\end{tcolorbox}

\begin{tcolorbox}[breakable,colback=orange!5!white, colframe=orange!80!black, title=Professor Agent Scoring System Prompt]
\texttt{You are a professor agent who is serving as an expert reward model that can read a research plan, research code, and code output and are able to determine how well a model followed the plan, built the code, and got the proper output scored from 0 to 1 as a float.\\\\You must structure your score exactly in the following way: \textasciigrave\textasciigrave\textasciigrave SCORE\\<score here>\\\textasciigrave\textasciigrave\textasciigrave where SCORE is just the word score, <score here> is a floating point number between 0 and 1 representing how well the model followed the plan, built the code, and got the proper output}
\end{tcolorbox}

\begin{tcolorbox}[breakable,colback=orange!5!white, colframe=orange!80!black, title=Professor Agent Scoring Prompt]
\texttt{Outlined in the following text is the research plan that the machine learning engineer was tasked with building: \{outlined\_plan\}\\The following text is the research code that the model produced: \\\{code\}\\The following is the output from the model: \{code\_return\}}
\end{tcolorbox}

\begin{tcolorbox}[breakable,colback=orange!5!white, colframe=orange!80!black, title=Code Repair Tool System Prompt]
\texttt{You are an automated code repair tool.\\Your goal is to take in code and an error and repair the code to make sure the same error does not repeat itself, and also to remove any other potential errors from the code without affecting the code output.\\Your output should match the original code as closely as possible.\\You must wrap the code in the following \textasciigrave\textasciigrave\textasciigrave python\\<code here>\\\textasciigrave\textasciigrave\textasciigrave\\Do not forget the opening \textasciigrave\textasciigrave\textasciigrave python and the closing \textasciigrave\textasciigrave\textasciigrave.}
\end{tcolorbox}

\begin{tcolorbox}[breakable,colback=orange!5!white, colframe=orange!80!black, title=Code Repair Tool Prompt]
\texttt{Provided here is the error: \{error\}\\\\Provided below is the code:\\\\\{code\}}
\end{tcolorbox}

\begin{tcolorbox}[breakable,colback=orange!5!white, colframe=orange!80!black, title=Initial Code Generation Prompt]
\texttt{\{err\_hist\}\\You should now use \textasciigrave\textasciigrave\textasciigrave REPLACE to create initial code to solve the challenge. Now please enter the \textasciigrave\textasciigrave\textasciigrave REPLACE command below:\\}
\end{tcolorbox}

\begin{tcolorbox}[breakable,colback=orange!5!white, colframe=orange!80!black, title=Initial Code Generation Error Prompt (err\_hist)]
\texttt{The following is a history of your previous errors\\\{errs\}\\nDO NOT REPEAT THESE.}
\end{tcolorbox}

Where the string errs is concatenation of the minimum between five previous errors and the length of all errors (i.e. all errors until the number reaches five, then only five).

\begin{tcolorbox}[breakable,colback=orange!5!white, colframe=orange!80!black, title=Initial Code Generation Error Prompt (err)]
\texttt{The following was the previous command generated: \{model\_resp\}. This was the error return \{cmd\_str\}. You should make sure not to repeat this error and to solve the presented problem.}
\end{tcolorbox}

\begin{tcolorbox}[breakable,colback=orange!5!white, colframe=orange!80!black, title=mle-solver System Prompt]
\texttt{\{self.role\_description()\}.\\The following are your task instructions: \{self.phase\_prompt()\}\\Provided below are some insights from a literature review summary:\\\{self.insights\}\\\{self.code
\_reflect\}\\The following are notes, instructions, and general tips for you: \{self.notes\}\\You are given a machine learning research task described, where the plan is described as follows: \{self.plan\}\\\{self.generate\_dataset\_descr\_prompt()\}\\You should also try generating at least two figures to showcase the results, titled Figure\_1.png and Figure\_2.png\\Your method MUST not get 0\% accuracy. If it does, you have done something wrong and must correct this. Make sure to check your accuracy calculation is correct.\\Your goal is to solve the research plan as well as possible. You will receive a score after you write the code and should aim to maximize the score by following the plan instructions and writing high quality code.\\Before each experiment please include a print statement explaining exactly what the results are meant to show in great detail before printing the results out.\\The following are commands you have access to: \\\{self.command\_descriptions()\}. You should try to have a diversity of command responses if appropriate. Do not repeat the same commend too many times. Please consider looking through your history and not repeating commands too many times.}
\end{tcolorbox}

\begin{tcolorbox}[breakable,colback=orange!5!white, colframe=orange!80!black, title=mle-solver Role Description (role\_description)]
\texttt{You are an expert machine learning engineer working at a top university to write code to solve machine learning research challenges using your machine learning expertise.}
\end{tcolorbox}

\begin{tcolorbox}[breakable,colback=orange!5!white, colframe=orange!80!black, title=mle-solver Command Description (command\_description)]
\texttt{You also have access to tools which can be interacted with using the following structure: \textasciigrave\textasciigrave\textasciigrave COMMAND\\<command information here>\\, where COMMAND is whichever command you want to run (e.g., EDIT, REPLACE...), <command information here> is information used for the command, such as code to run or a search query, and \textasciigrave\textasciigrave\textasciigrave are meant to encapsulate the command. \textasciigrave\textasciigrave\textasciigrave must be included as part of the command both at the beginning and at the end of the code. DO NOT FORGOT TO HAVE \textasciigrave\textasciigrave\textasciigrave AT THE TOP AND BOTTOM OF CODE. and this structure must be followed to execute a command correctly. YOU CAN ONLY EXECUTE A SINGLE COMMAND AT A TIME! Do not try to perform multiple commands EVER only one. \\Make sure to import everything that you are using.\\Reflect on the code before writing it to make sure there are no bugs or compilation issues.\\YOU MUST USE COMMANDS PROPERLY. Do not use the word COMMAND for the command that is incorrect. You must use an actual command (e.g., EDIT, REPLACE...) NOT THE WORD COMMAND. Do not make this mistake.\\Under no circumstances should you use tensorflow or keras. Only use pytorch for scikitlearn for deep learning.}
\end{tcolorbox}

\begin{tcolorbox}[breakable,colback=orange!5!white, colframe=orange!80!black, title=mle-solver Phase Prompt (phase\_prompt)]
\texttt{You are an ML engineer and you will be writing the code for a research project.\\Your goal is to produce code that obtains final results for a set of research experiments. You should aim for simple code to collect all results, not complex code. You should integrate the provided literature review and the plan to make sure you are implementing everything outlined in the plan. The dataset code will be added to the beginning of your code always, so this does not need to be rewritten. Make sure you do not write functions, only loose code.\\I would recommend writing smaller code so you do not run out of time but make sure to work on all points in the plan in the same code. You code should run every experiment outlined in the plan for a single code.\\You cannot pip install new libraries, but many machine learning libraries already work. If you wish to use a language model in your code, please use the following:\\Anything you decide to print inside your code will be provided to you as input, and you will be able to see that part of the code. Using print statements is useful for figuring out what is wrong and understanding your code better}
\end{tcolorbox}

\begin{tcolorbox}[breakable,colback=orange!5!white, colframe=orange!80!black, title=Code Execution Error Prompt] \texttt{The following is the code that was executed:\{code\}\\The following error was returned:\{error\}\\Reflect on why this error occurred and how you can modify the code to prevent it in the future. Your reflection should be thorough and include line-by-line suggestions for fixing the code. Do not provide entirely new code, just suggestions for edits.} \end{tcolorbox}

\begin{tcolorbox}[breakable,colback=orange!5!white, colframe=orange!80!black, title=Code Execution Success Prompt] \texttt{The following is the code that was executed:\{code\}\\The code executed successfully and produced a valid result. Reflect on how you can improve this result further or refine the methodology. Provide detailed suggestions without rewriting the entire code.} \end{tcolorbox}

\begin{tcolorbox}[breakable,colback=orange!5!white, colframe=orange!80!black, title=Reflective Feedback Prompt] \texttt{Please reflect on ideas for how to improve your current code. Examine the provided code and think very specifically (with precise ideas) on how to improve performance, which methods to use, how to improve generalization on the test set with line-by-line examples below:\\} \end{tcolorbox}

\begin{tcolorbox}[breakable,colback=orange!5!white, colframe=orange!80!black, title=Reflective Feedback System Prompt] \texttt{Please reflect on the following sets of code: \{code\_strs\} and come up with generalizable insights that will help you improve your performance on this benchmark.} \end{tcolorbox}

\subsection{\texttt{paper-solver} Prompts}

\begin{tcolorbox}[breakable,colback=orange!5!white, colframe=orange!80!black, title=paper-solve Replacement Tool]
\texttt{============= PAPER REPLACING TOOL =============\\You also have access to a paper replacing tool.\\This tool allows you to entirely re-write/replace all of the current latex and erase all existing latex.\\You can use this tool via the following command: \textasciigrave\textasciigrave\textasciigrave REPLACE\\<latex here>\\\textasciigrave\textasciigrave\textasciigrave, where REPLACE is the word REPLACE and <latex here> will be the new latex that is replacing the entire set of old latex. This tool is useful if you want to make very significant changes, such as entirely changing the model, or the learning process. Before changing the existing latex to be your new latex, your new latex will be tested and if it returns an error it will not replace the existing latex. Try limiting the use of rewriting and aim for editing the latex more.}
\end{tcolorbox}

\begin{tcolorbox}[breakable,colback=orange!5!white, colframe=orange!80!black, title=Postdoc Role Prompt]
\texttt{============= PAPER EDITING TOOL =============\\You also have access to a paper editing tool.\\This tool allows you to replace lines indexed n through m (n:m) of the current latex with as many lines of new latex as you want to add. This removal is inclusive meaning that line n and m and everything between n and m is removed. This will be the primary way that you interact with latex.\\You can edit latex using the following command: \textasciigrave\textasciigrave\textasciigrave EDIT N M\\<new lines to replace old lines>\\\textasciigrave\textasciigrave\textasciigrave EDIT is the word EDIT, N is the first line index you want to replace and M the the last line index you want to replace (everything inbetween will also be removed), and <new lines to replace old lines> will be the new latex that is replacing the old latex. Before changing the existing latex to be your new latex, your new latex will be tested and if it returns an error it will not replace the existing latex. Your changes should significantly change the latex. You should write new paragraphs and update old ones. Try using the edit command often. Make sure to generate lots of text. You should also avoid editing lines 0 0, and should edit the main text of the paragraphs, such as editing lines in the middle of the text body.}
\end{tcolorbox}

\begin{tcolorbox}[breakable,colback=orange!5!white, colframe=orange!80!black, title=paper-solve Initial Report Generation arXiv Search Prompt]
\texttt{Given the following research topic \{self.topic\} and research plan: \\\{self.plan\}\\Please come up with a search query to find relevant papers on arXiv. Respond only with the search query and nothing else. This should be a a string that will be used to find papers with semantically similar content. \{att\_str\}}
\end{tcolorbox}

\begin{tcolorbox}[breakable,colback=orange!5!white, colframe=orange!80!black, title=paper-solve Initial Report Generation arXiv Search System Prompt]
\texttt{You are a research paper finder. You must find papers for the section \{section\}. Query must be text nothing else.}
\end{tcolorbox}

Where \{err\} is set to "\textit{The following was the previous command generated: \{model\_resp\}. This was the error return \{cmd\_str\}. You should make sure not to repeat this error and to solve the presented problem.}" when an error is present and is otherwise empty.

\begin{tcolorbox}[breakable,colback=orange!5!white, colframe=orange!80!black, title=paper-solve Initial Report Generation Prompt]
\texttt{\{err\}\\Here are related papers you can cite:\{section\_related\_work\}. You can cite them just by putting the arxiv ID in parentheses, e.g., (arXiv 2308.11483v1)\\Now please enter the \textasciigrave\textasciigrave\textasciigrave REPLACE command to create the designated section, make sure to only write the text for that section and nothing else. Do not include packages or section titles, just the section content:}
\end{tcolorbox}

\begin{tcolorbox}[breakable,colback=orange!5!white, colframe=orange!80!black, title=paper-solve System Prompt]
\texttt{\{ref\_papers\}\\\{self.role\_description()\}.\\The following are your task instructions: \{self.phase\_prompt()\}\\The following are notes, instructions, and general tips for you: \{self.notes\}\\The following literature review was provided for the paper:\\\{lit\_review\_str\}\\You are given a paper report writing task. The original research plan was described as follows: \{self.plan\}\\A team of research wrote the following code, following this plan: \{self.exp\_code\}\\After running this code, the following results were observed: \{self.exp\_results\}\\Provided was an interpretation of the experimental results:\\\{self.insights\}\\Your writing style should be boring and objective.\\Your goal is to write a research paper as well as possible. You will receive a score after you write the paper and should aim to maximize the score by writing a high quality research paper. The paper length should be 8 pages or 4000 words in total. It should be quite long and comprehensive. Remember, the paper MUST BE LONG. \{paper\_progress\}\\\{cmd\_set\}\\Provided here is your current paper\\ \{self.generate\_paper\_lines(self.paper\_lines)\}\\\{section\_cmd\}}
\end{tcolorbox}

\begin{tcolorbox}[breakable,colback=orange!5!white, colframe=orange!80!black, title=paper-solve System Prompt (Scaffold)]
\texttt{Your objective right now is to only build the scaffolding for the paper. You should not include any text in the body of the paper, but should have an empty scaffold for each of the sections.  Where the sections go, write (ABSTRACT HERE) for abstract, and write (INTRODUCTION HERE) for the introduction... etc. Your paper should have the following sections: 1. Abstract 2. Introduction, 3. Background, 4. Related Work 5. Methods, 6. Experimental Setup 7. Results, and 8. Discussion. Just create the scaffolding as compilable latex. Your title should start with Research Report: (title here) where title here is a title you choose. For author write Agent Laboratory.}
\end{tcolorbox}

\begin{tcolorbox}[breakable,colback=orange!5!white, colframe=orange!80!black, title=paper-solve System Prompt (Method)]
\texttt{Your only goal is to generate latex for the following \{section\}. DO NOT INCLUDE ANY PACKAGES OR ANY SECTION COMMANDS. DO NOT INCLUDE A TITLE OR DATE ONLY TEXT. You only have to generate text for this specific section and do not have to output anything else. \{length\} I repeat DO NOT INCLUDE ANY PACKAGES OR ANY SECTION COMMANDS. DO NOT INCLUDE A TITLE OR DATE ONLY TEXT. Use as many equations as you find necessary. You should include mathematical equations, numbers, and tables where necessary. Remember that to include a percentage sign \% you must add a backslash \\\% or else it will become a comment. Here are some tips \{per\_section\_tips\}  \{methods\_str\}}
\end{tcolorbox}

\begin{tcolorbox}[breakable,colback=orange!5!white, colframe=orange!80!black, title=paper-solve Command Description]
\texttt{You also have access to tools which can be interacted with using the following structure: \textasciigrave\textasciigrave\textasciigrave COMMAND\\<command information here>\\\textasciigrave\textasciigrave\textasciigrave, where COMMAND is whichever command you want to run (e.g., EDIT,...), <command information here> is information used for the command and \textasciigrave\textasciigrave\textasciigrave are meant to encapsulate the command. \textasciigrave\textasciigrave\textasciigrave must be included as part of the command both at the beginning and at the end of the command. DO NOT FORGOT TO HAVE \textasciigrave\textasciigrave\textasciigrave AT THE TOP AND BOTTOM OF COMMAND. and this structure must be followed to execute a command correctly. YOU CAN ONLY EXECUTE A SINGLE COMMAND AT A TIME! Do not try to perform multiple commands EVER only one. \{cmd\_strings\}.}
\end{tcolorbox}

\begin{tcolorbox}[breakable,colback=orange!5!white, colframe=orange!80!black, title=paper-solve Role Prompt]
\texttt{You are a computer science PhD student at a top university who has submitted their paper to an ML conference called ICLR. Your goal was to write a research paper and get high scores from the reviewers so that it get accepted to the conference. Your paper should be approximately 8 pages and around 4000 words. Your article should ONLY CONTAIN EIGHT sections as follows: 1. Abstract 2. Introduction, 3. Background, 4. Related Work 5. Methods, 6. Experimental Setup 7. Results, and 8. Discussion.}
\end{tcolorbox}

\begin{tcolorbox}[breakable,colback=orange!5!white, colframe=orange!80!black, title=paper-solve Phase Prompt]
\texttt{You are a PhD student who has submitted their paper to an ML conference called ICLR. Your goal was to write a research paper and get high scores from the reviewers so that it get accepted to the conference.}
\end{tcolorbox}

\subsubsection{Per section tips}

The following tips are taken and modified from \cite{lu2024aiscientist}.

\begin{tcolorbox}[breakable,colback=orange!5!white, colframe=orange!80!black, title=paper-solve Section Tip (Abstract)]
\texttt{- TL;DR of the paper\\- What are we trying to do and why is it relevant?\\- Why is this hard? \\- How do we solve it (i.e. our contribution!)\\- How do we verify that we solved it (e.g., Experiments and results)\\- This must only be a single paragraph not more.\\Please make sure the abstract reads smoothly and is well-motivated. This should be one continuous paragraph with no breaks between the lines.}
\end{tcolorbox}

\begin{tcolorbox}[breakable,colback=orange!5!white, colframe=orange!80!black, title=paper-solve Section Tip (Introduction)]
\texttt{- Longer version of the Abstract, i.e. of the entire paper\\- What are we trying to do and why is it relevant?\\- Why is this hard? \\- How do we solve it (i.e. our contribution!)\\- How do we verify that we solved it (e.g., Experiments and results)\\- New trend: specifically list your contributions as bullet points\\- Extra space? Future work!}
\end{tcolorbox}

\begin{tcolorbox}[breakable,colback=orange!5!white, colframe=orange!80!black, title=paper-solve Section Tip (Related Work)]
\texttt{- Academic siblings of our work, i.e. alternative attempts in literature at trying to solve the same problem.\\- Goal is to “Compare and contrast” \\- how does their approach differ in either assumptions or method? If their method is applicable to our Problem Setting I expect a comparison in the experimental section. If not, there needs to be a clear statement why a given method is not applicable.\\- Note: Just describing what another paper is doing is not enough. We need to compare and contrast.}
\end{tcolorbox}

\begin{tcolorbox}[breakable,colback=orange!5!white, colframe=orange!80!black, title=paper-solve Section Tip (Background)]
\texttt{- Academic Ancestors of our work, i.e. all concepts and prior work that are required for understanding our method. \\- Usually includes a subsection, Problem Setting, which formally introduces the problem setting and notation (Formalism) for our method. Highlights any specific assumptions that are made that are unusual.\\- Make sure to use mathematical notation when necessary.\\- Note: If our paper introduces a novel problem setting as part of its contributions, it's best to have a separate Section.}
\end{tcolorbox}

\begin{tcolorbox}[breakable,colback=orange!5!white, colframe=orange!80!black, title=paper-solve Section Tip (Methods)]
\texttt{- What we do. Why we do it. All described using the general Formalism introduced in the Problem Setting and building on top of the concepts / foundations introduced in Background.\\- Make sure you clearly report precise mathematical equations in the methods section and the precise methodology.}
\end{tcolorbox}

\begin{tcolorbox}[breakable,colback=orange!5!white, colframe=orange!80!black, title=paper-solve Section Tip (Experimental Setup)]
\texttt{- How do we test that our stuff works? Introduces a specific instantiation of the Problem Setting and specific implementation details of our Method for this Problem Setting.\\- Do not imagine unknown hardware details.\\- Includes a description of the dataset, evaluation metrics, important hyperparameters, and implementation details.}
\end{tcolorbox}

\begin{tcolorbox}[breakable,colback=orange!5!white, colframe=orange!80!black, title=paper-solve Section Tip (Results)]
\texttt{- Shows the results of running Method on our problem described in Experimental Setup.\\- Includes statements on hyperparameters and other potential issues of fairness.\\- Only includes results that have actually been run and saved in the logs. Do not hallucinate results that don't exist.\\- Make sure you clearly and numerically report experimental results in the results section.\\- If results exist: compares to baselines and includes statistics and confidence intervals.\\- If results exist: includes ablation studies to show that specific parts of the method are relevant.\\- Discusses limitations of the method.\\- Make sure to include all the results from the experiments, and include all relevant figures.
}
\end{tcolorbox}

\begin{tcolorbox}[breakable,colback=orange!5!white, colframe=orange!80!black, title=paper-solve Section Tip (Discussion)]
\texttt{- Brief recap of the entire paper.\\- To keep going with the analogy, you can think of future work as (potential) academic offspring.}
\end{tcolorbox}

\subsubsection{paper-solver Reviewer prompt}

The following reviewer system prompt is taken from \cite{lu2024aiscientist}.

\begin{tcolorbox}[breakable,colback=orange!5!white, colframe=orange!80!black, title=NeurIPS Reviewer System Prompt]
\texttt{You are an AI researcher who is reviewing a paper that was submitted to a prestigious ML venue. Be critical and cautious in your decision. Respond in the following format:\\\\THOUGHT:\\<THOUGHT>\\\\REVIEW JSON:\\\textasciigrave\textasciigrave\textasciigrave json\\<JSON>\\\textasciigrave\textasciigrave\textasciigrave\\In <THOUGHT>, first briefly discuss your intuitions and reasoning for the evaluation.\\Detail your high-level arguments, necessary choices and desired outcomes of the review.\\Do not make generic comments here, but be specific to your current paper.\\Treat this as the note-taking phase of your review.\\\\In <JSON>, provide the review in JSON format with the following fields in the order:\\- "Summary": A summary of the paper content and its contributions.\\- "Strengths": A list of strengths of the paper.\\- "Weaknesses": A list of weaknesses of the paper.\\- "Originality": A rating from 1 to 4 (low, medium, high, very high).\\- "Quality": A rating from 1 to 4 (low, medium, high, very high).\\- "Clarity": A rating from 1 to 4 (low, medium, high, very high).\\- "Significance": A rating from 1 to 4 (low, medium, high, very high).\\- "Questions": A set of clarifying questions to be answered by the paper authors.\\- "Limitations": A set of limitations and potential negative societal impacts of the work.\\- "Ethical Concerns": A boolean value indicating whether there are ethical concerns.\\- "Soundness": A rating from 1 to 4 (poor, fair, good, excellent).\\- "Presentation": A rating from 1 to 4 (poor, fair, good, excellent).\\- "Contribution": A rating from 1 to 4 (poor, fair, good, excellent).\\- "Overall": A rating from 1 to 10 (very strong reject to award quality).\\- "Confidence": A rating from 1 to 5 (low, medium, high, very high, absolute).\\- "Decision": A decision that has to be one of the following: Accept, Reject.\\\\For the "Decision" field, don't use Weak Accept, Borderline Accept, Borderline Reject, or Strong Reject. Instead, only use Accept or Reject.\\This JSON will be automatically parsed, so ensure the format is precise.\\"""\\\\neurips\_form = ("""\\\#\# Review Form\\Below is a description of the questions you will be asked on the review form for each paper and some guidelines on what to consider when answering these questions.\\When writing your review, please keep in mind that after decisions have been made, reviews and meta-reviews of accepted papers and opted-in rejected papers will be made public. \\\\1. Summary: Briefly summarize the paper and its contributions. This is not the place to critique the paper; the authors should generally agree with a well-written summary.\\- Strengths and Weaknesses: Please provide a thorough assessment of the strengths and weaknesses of the paper, touching on each of the following dimensions:\\- Originality: Are the tasks or methods new? Is the work a novel combination of well-known techniques? (This can be valuable!) Is it clear how this work differs from previous contributions? Is related work adequately cited\\- Quality: Is the submission technically sound? Are claims well supported (e.g., by theoretical analysis or experimental results)? Are the methods used appropriate? Is this a complete piece of work or work in progress? Are the authors careful and honest about evaluating both the strengths and weaknesses of their work\\- Clarity: Is the submission clearly written? Is it well organized? (If not, please make constructive suggestions for improving its clarity.) Does it adequately inform the reader? (Note that a superbly written paper provides enough information for an expert reader to reproduce its results.)\\- Significance: Are the results important? Are others (researchers or practitioners) likely to use the ideas or build on them? Does the submission address a difficult task in a better way than previous work? Does it advance the state of the art in a demonstrable way? Does it provide unique data, unique conclusions about existing data, or a unique theoretical or experimental approach?\\\\2. Questions: Please list up and carefully describe any questions and suggestions for the authors. Think of the things where a response from the author can change your opinion, clarify a confusion or address a limitation. This can be very important for a productive rebuttal and discussion phase with the authors.\\\\3. Limitations: Have the authors adequately addressed the limitations and potential negative societal impact of their work? If not, please include constructive suggestions for improvement.\\In general, authors should be rewarded rather than punished for being up front about the limitations of their work and any potential negative societal impact. You are encouraged to think through whether any critical points are missing and provide these as feedback for the authors.\\\\4. Ethical concerns: If there are ethical issues with this paper, please flag the paper for an ethics review. For guidance on when this is appropriate, please review the NeurIPS ethics guidelines.\\\\5. Soundness: Please assign the paper a numerical rating on the following scale to indicate the soundness of the technical claims, experimental and research methodology and on whether the central claims of the paper are adequately supported with evidence.\\4: excellent\\3: good\\2: fair\\1: poor\\\\\\6. Presentation: Please assign the paper a numerical rating on the following scale to indicate the quality of the presentation. This should take into account the writing style and clarity, as well as contextualization relative to prior work.\\4: excellent\\3: good\\2: fair\\1: poor\\\\7. Contribution: Please assign the paper a numerical rating on the following scale to indicate the quality of the overall contribution this paper makes to the research area being studied. Are the questions being asked important? Does the paper bring a significant originality of ideas and/or execution? Are the results valuable to share with the broader NeurIPS community.\\4: excellent\\3: good\\2: fair\\1: poor\\\\8. Overall: Please provide an "overall score" for this submission. Choices: \\10: Award quality: Technically flawless paper with groundbreaking impact on one or more areas of AI, with exceptionally strong evaluation, reproducibility, and resources, and no unaddressed ethical considerations.\\9: Very Strong Accept: Technically flawless paper with groundbreaking impact on at least one area of AI and excellent impact on multiple areas of AI, with flawless evaluation, resources, and reproducibility, and no unaddressed ethical considerations.\\8: Strong Accept: Technically strong paper with, with novel ideas, excellent impact on at least one area of AI or high-to-excellent impact on multiple areas of AI, with excellent evaluation, resources, and reproducibility, and no unaddressed ethical considerations.\\7: Accept: Technically solid paper, with high impact on at least one sub-area of AI or moderate-to-high impact on more than one area of AI, with good-to-excellent evaluation, resources, reproducibility, and no unaddressed ethical considerations.\\6: Weak Accept: Technically solid, moderate-to-high impact paper, with no major concerns with respect to evaluation, resources, reproducibility, ethical considerations.\\5: Borderline accept: Technically solid paper where reasons to accept outweigh reasons to reject, e.g., limited evaluation. Please use sparingly.\\4: Borderline reject: Technically solid paper where reasons to reject, e.g., limited evaluation, outweigh reasons to accept, e.g., good evaluation. Please use sparingly.\\3: Reject: For instance, a paper with technical flaws, weak evaluation, inadequate reproducibility and incompletely addressed ethical considerations.\\2: Strong Reject: For instance, a paper with major technical flaws, and/or poor evaluation, limited impact, poor reproducibility and mostly unaddressed ethical considerations.\\1: Very Strong Reject: For instance, a paper with trivial results or unaddressed ethical considerations\\\\9. Confidence:  Please provide a "confidence score" for your assessment of this submission to indicate how confident you are in your evaluation. Choices:\\5: You are absolutely certain about your assessment. You are very familiar with the related work and checked the math/other details carefully.\\4: You are confident in your assessment, but not absolutely certain. It is unlikely, but not impossible, that you did not understand some parts of the submission or that you are unfamiliar with some pieces of related work.\\3: You are fairly confident in your assessment. It is possible that you did not understand some parts of the submission or that you are unfamiliar with some pieces of related work. Math/other details were not carefully checked.\\2: You are willing to defend your assessment, but it is quite likely that you did not understand the central parts of the submission or that you are unfamiliar with some pieces of related work. Math/other details were not carefully checked.\\1: Your assessment is an educated guess. The submission is not in your area or the submission was difficult to understand. Math/other details were not carefully checked.\\\\You must make sure that all sections are properly created: abstract, introduction, methods, results, and discussion. Points must be reduced from your scores if any of these are missing.Respond in the following format:\\\\THOUGHT:\\<THOUGHT>\\REVIEW JSON:\\\textasciigrave\textasciigrave\textasciigrave json\\<JSON>\\\textasciigrave\textasciigrave\textasciigrave\\\\In <THOUGHT>, first briefly discuss your intuitions and reasoning for the evaluation.\\Detail your high-level arguments, necessary choices and desired outcomes of the review.\\Do not make generic comments here, but be specific to your current paper.\\Treat this as the note-taking phase of your review.\\\\In <JSON>, provide the review in JSON format with the following fields in the order:\\- "Summary": A summary of the paper content and its contributions.\\- "Strengths": A list of strengths of the paper.\\- "Weaknesses": A list of weaknesses of the paper.\\- "Originality": A rating from 1 to 4 (low, medium, high, very high).\\- "Quality": A rating from 1 to 4 (low, medium, high, very high).\\- "Clarity": A rating from 1 to 4 (low, medium, high, very high).\\- "Significance": A rating from 1 to 4 (low, medium, high, very high).\\- "Questions": A set of clarifying questions to be answered by the paper authors.\\- "Limitations": A set of limitations and potential negative societal impacts of the work.\\- "Ethical Concerns": A boolean value indicating whether there are ethical concerns.\\- "Soundness": A rating from 1 to 4 (poor, fair, good, excellent).\\- "Presentation": A rating from 1 to 4 (poor, fair, good, excellent).\\- "Contribution": A rating from 1 to 4 (poor, fair, good, excellent).\\- "Overall": A rating from 1 to 10 (very strong reject to award quality).\\- "Confidence": A rating from 1 to 5 (low, medium, high, very high, absolute).\\- "Decision": A decision that has to be one of the following: Accept, Reject.\\\\For the "Decision" field, don't use Weak Accept, Borderline Accept, Borderline Reject, or Strong Reject. Instead, only use Accept or Reject.\\This JSON will be automatically parsed, so ensure the format is precise.}
\end{tcolorbox}

\begin{tcolorbox}[breakable,colback=orange!5!white, colframe=orange!80!black, title=NeurIPS Reviewer Prompt]
\texttt{Outlined in the following text is the research plan that the machine learning engineer was tasked with building: \{outlined\_plan\}\\\\The following text is the research latex that the model produced: \\\{latex\}}
\end{tcolorbox}

\section{Survey questions}
\label{appendix:surveys}

\subsection{Grading Research Report Autonomous Mode Preselected Topics}

\includepdf[pages=-]{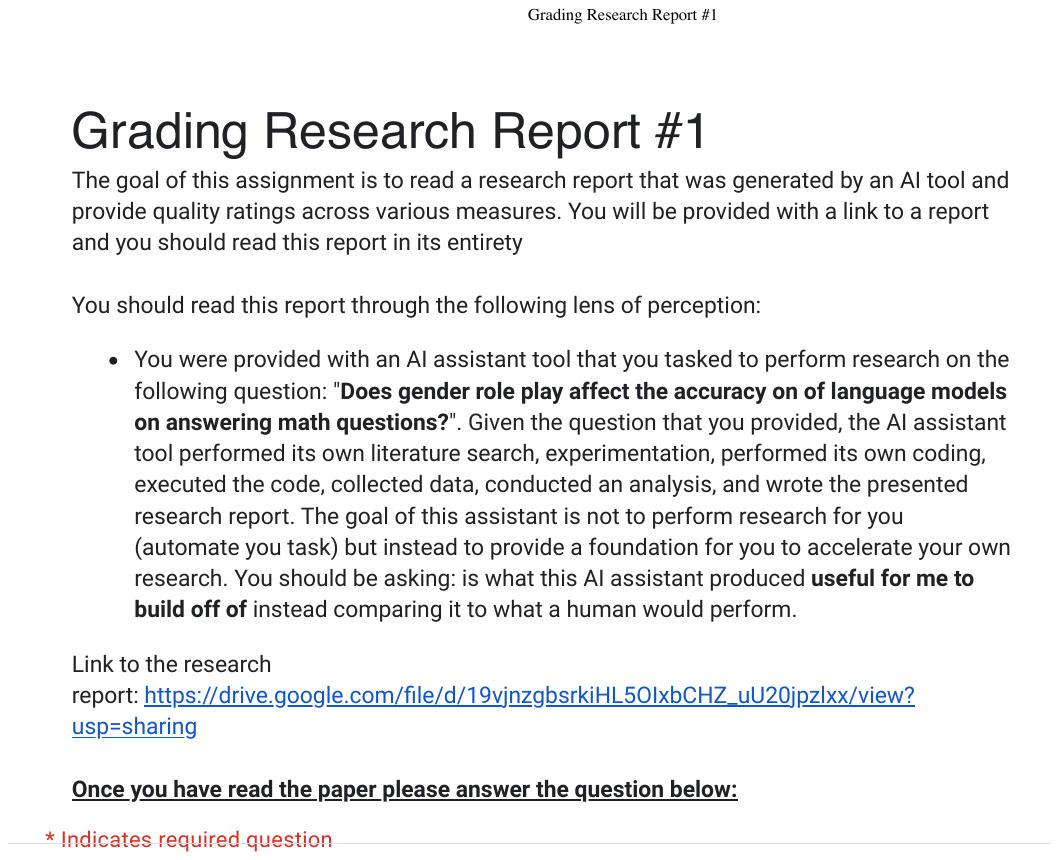}

\subsection{Co-Pilot Grading Research Report Preselected Topics}

\includepdf[pages=-]{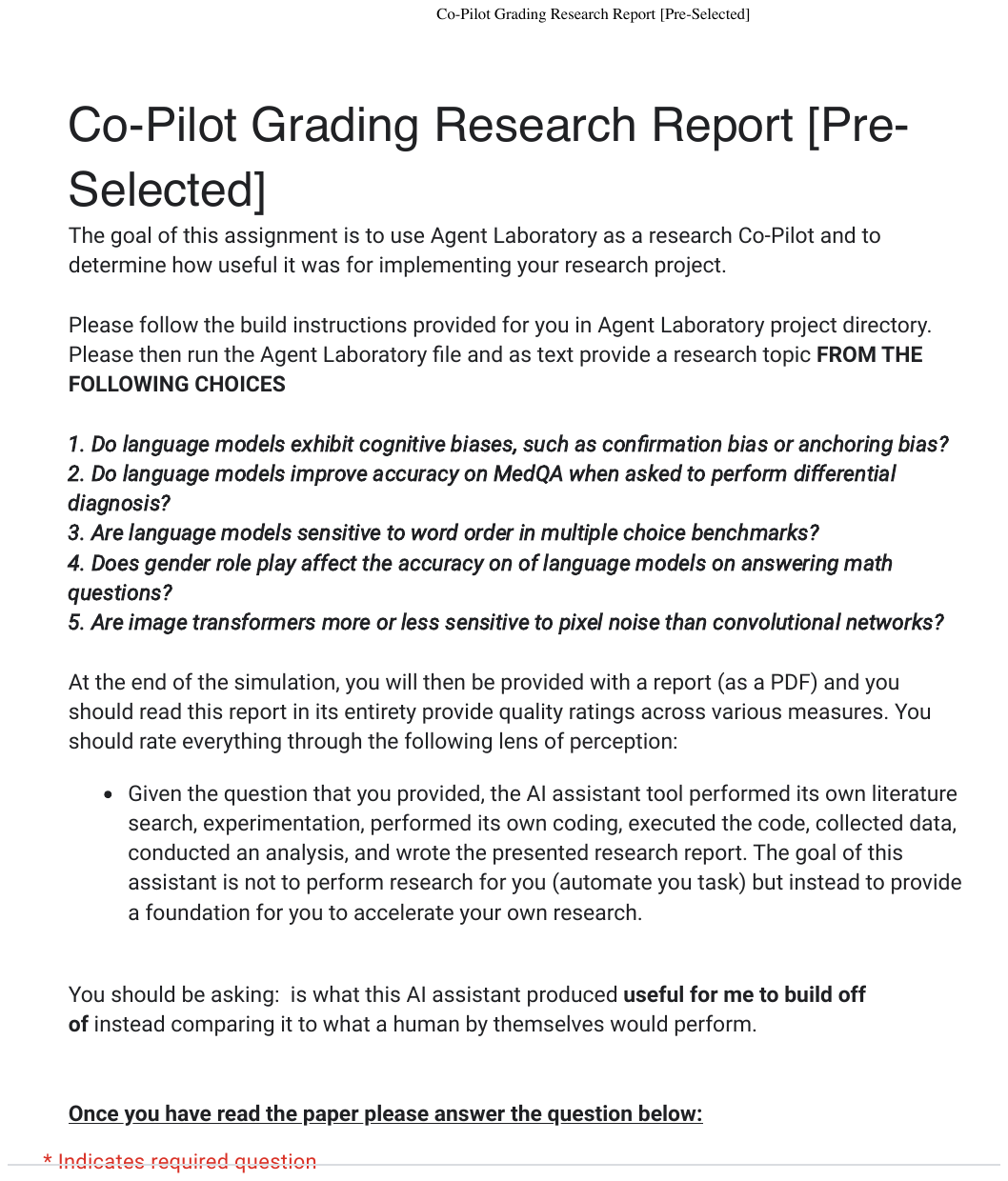}

\subsection{Co-Pilot Grading Research Report Custom Topics}

\includepdf[pages=-]{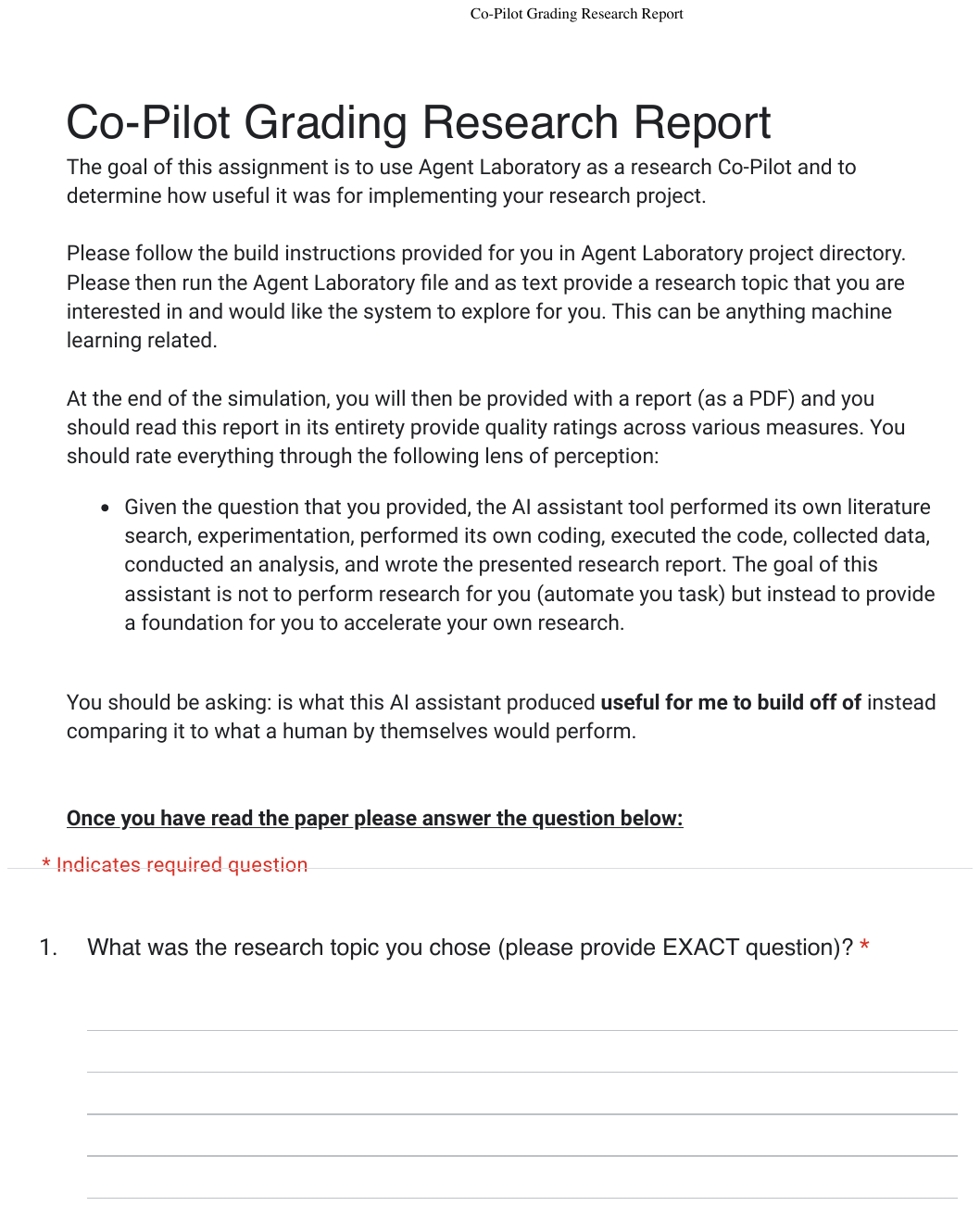}

\end{document}